\documentclass[journal,9pt,table]{IEEEtran}

\usepackage[style=ACM-Reference-Format, backend=bibtex, sorting=none, maxbibnames=2]{biblatex}

\newcommand\timesdiv{\mathbin{\vcenter{\hbox{%
   $\begin{array}{@{}c@{}}\times\\[-1.3ex]\div\end{array}$}}}}

\usepackage{amsmath,amssymb,amsfonts}
\usepackage{graphicx}
\usepackage{adjustbox} 
\usepackage{textcomp}

\newcommand{\blank}[1]{\hspace*{#1}} 
\usepackage[table]{xcolor}
\definecolor{lightgreen}{rgb}{0.5, 0.95, 0.75}
\definecolor{lightgray}{gray}{0.9}
\definecolor{LightCyan}{rgb}{0.88,1,1}
\definecolor{green}{rgb}{0.45,0.85,0.65}
\definecolor{dark_green}{rgb}{0, 0.4, 0.1}
\definecolor{sam_color}{rgb}{1.0,0.3,0.2}
\definecolor{dark_blue}{rgb}{0.05,0.05,0.89}
\definecolor{lighter_green}{rgb}{0.8, 1, 0.8}

\newcolumntype{a}{>{\columncolor{lightgray}}c}
\newcolumntype{b}{>{\columncolor{white}}c}

\usepackage{tikz} 

\usepackage{placeins}
\usepackage{varioref}
\usepackage{eqparbox}
\usepackage{footnote}
\makesavenoteenv{tabular}
\makesavenoteenv{table}
\usepackage{url}
\usepackage{booktabs}
\usepackage{multirow}
\usepackage{multicol}
\usepackage{array}
\newcolumntype{?}[1]{!{\vrule width #1}}

 \usepackage{tabularx}
\usepackage{bm}
\usepackage{soul}
\usepackage{tikz}
\usepackage{wrapfig}
\usepackage{amssymb}
\usepackage[bottom]{footmisc}

\usepackage{enumitem}
\usepackage{float}
\floatstyle{plaintop}
\restylefloat{table}
\usepackage[tableposition=top]{caption} 
\usepackage{multirow, booktabs}
\usepackage{siunitx}

\usepackage{tabu,booktabs}
\usepackage{makecell}

\usepackage{hyperref}
\hypersetup{
    colorlinks,
    linkcolor={red!50!black},
    citecolor={green!40!black},
    urlcolor={blue!80!black}
}

\usepackage{threeparttable}	
\usepackage{threeparttablex}
\usepackage{pifont}
\newcommand{\cmark}{\ding{51}}%
\newcommand{\xmark}{\ding{55}}%
\usepackage{tabularx}
\usepackage{hhline}
\usepackage{textcomp}
\usepackage[linesnumbered,ruled,vlined]{algorithm2e}
\usepackage{tabu}
\usepackage{siunitx}
\usepackage{amsmath, mathtools}
\usepackage{subcaption}
\usepackage{lipsum}

\usepackage{wrapfig} 

\newcolumntype{L}{>{\centering\arraybackslash}m{1.2cm}}
\usetikzlibrary{arrows,fit,positioning}

\usetikzlibrary{calc}
\usetikzlibrary{decorations.pathmorphing}
\newcolumntype{C}[1]{>{\centering\let\newline\\\arraybackslash\hspace{0pt}}m{#1}}
\newcolumntype{C}{>{\centering\arraybackslash}p{2.5em}}

\newcommand{\figbox}[1]{%
  \fbox{%
    \vbox to 1in{%
    \vfil
    \hbox to 2in{%
      \hfil
      #1%
      \hfil}%
    \vfil}}}

\newcommand{\pname}{RAPID}

\begin{document}
\bstctlcite{IEEEexample:BSTcontrol}    

%
\title{RAPID: App\underline{R}oxim\underline{A}te \underline{P}ipelined Soft Mult\underline{I}pliers and \underline{D}ividers for High-Throughput and Energy-Efficiency
}

\author{\Large Zahra Ebrahimi, Muhammad Zaid, Mark Wijtvliet, Akash Kumar, \textit{Senior Member, IEEE} \vspace{-0.4cm}}
\maketitle



%

%
\begin{abstract}
The rapid updates in error-resilient applications along with their quest for high throughput
has motivated designing fast
approximate functional units for Field-Programmable Gate Arrays (FPGAs). 
Studies have proposed various imprecise functional techniques, albeit posed with three shortcomings: first, most existing inexact multipliers and dividers are specialized for Application-Specific Integrated Circuit (ASIC) platforms.
Therefore, due to the architectural differences of underlying building blocks in FPGA and ASIC, ASIC-customized designs have not yielded comparable improvements when directly synthesized and ported to FPGAs.
Second, state-of-the-art (SoA) approximate units are substituted, mostly in a single kernel of a multi-kernel application. Moreover, the \textit{end-to-end} assessment is adopted on the Quality of Results (QoR), but not on the \textit{overall} gained performance.
Finally, existing imprecise components are not
designed to support a pipelined approach, which could boost the operating frequency/throughput of, e.g., division-included applications.
In this paper, we propose~\pname, the first pipelined approximate multiplier and divider architectures, customized for FPGAs. The proposed units efficiently utilize
6-input Look-up Tables (6-LUTs) and fast carry chains to implement Mitchell's approximate algorithms. Our novel error-refinement scheme not only has negligible overhead over the baseline Mitchell’s approach, but also boosts its accuracy to 99.4\% for arbitrary size of multiplication and division.

Experimental results obtained with Xilinx Vivado demonstrate the efficiency of the proposed pipelined and non-pipelined \pname~multipliers and dividers over accurate counterparts. In particular,  4-stage pipelined architecture of 32-bit \pname~multiplier (divider) enables 3.3$\times$ (5.1$\times$) higher throughput, 2.3$\times$ (6.8$\times$) higher throughput/Watt, and 52\% (31\%) savings of LUTs, over their 4-stage pipelined, accurate IP counterparts.
Moreover, the end-to-end evaluations of non-pipelined \pname, deployed in three multi-kernel applications in the domains of bio-signal processing, image processing, and moving object tracking for Unmanned Air Vehicles (UAV) indicate up to
35\%, 33\%, and 45\% improvements in area, latency, and Area-Delay-Product (ADP), respectively, over accurate kernels, with negligible loss in QoR.
To springboard future research in reconfigurable and approximate computing communities, our implementations will be available and open-sourced at \href{https://cfaed.tu-dresden.de/pd-downloads}{\textcolor{blue}{\textit{https://cfaed.tu-dresden.de/pd-downloads}}}.

\end{abstract}

\begin{IEEEkeywords}
Field-Programmable Gate Arrays, Approximate Computing, Pipeline, Multiplier, Divider, Mitchell's Algorithm, Bio-signal Processing,
Unmanned Air Vehicles, High-Throughput, Energy-Efficiency.
\vspace{-0.2cm}
\end{IEEEkeywords}

 \section{ Introduction} \label{introduction}
The ever-growing demand for edge computing has become pronounced in the Internet of Things (IoT) era for a wide domain of applications, from bio-signal to various cutting-edge image processing. For example, wearable 24/7 health monitoring gadgets are becoming ubiquitous, especially noting that 47\% of cardiac diseases -- the main cause of death, worldwide -- occur outside of hospitals \cite{cvd_who, hospital_death}. Unmanned Aerial Vehicles (UAVs) such as drones are also proliferating for e.g., object/self tracking, search and surveillance, agricultural operations, and entertainment. 
Although Application-Specific Integrated Circuits (ASICs) are highly power-efficient for implementing the above-mentioned programs, off-the-shelf Field-Programmable Gate Arrays (FPGAs) have shown to serve as commercially-viable options owing to their rapid prototyping and post-fabrication datapath versatility
which can address the outpacing speed of algorithmic updates over the updates in hardware \cite{8977837, 10.1145/2821508, chatterjee2015real}. For example, the hardware of health gadgets should be able to adapt with various patients' physiological traits and the changes in heart’s activity.
High-throughput and/or energy-efficiency are also of high desire for the acceleration of such parallelizable applications that are repeatedly fed with a bulk of data.

FPGAs
are equipped with hard-wired DSP blocks to speed-up
multiplication, being the commonly-used function in bio-signal or image processing workloads. However, exploiting DSP blocks may not fulfill design requirements due to three reasons: first, their limited ratio versus Look-up Tables (LUTs), i.e., $<$0.001 could be insufficient for concurrent or multiplication-intensive applications.
Second, their fixed locations in FPGAs impose routing overhead and may result in reduced performance for some applications \cite{9344673, 10.1145/3195970.3195996, 4068926}. 
Finally, DSPs are unable to be efficiently-utilized for multiplication with precision, smaller than 18$\times$18 bit \cite{8532582, 8332524}. Therefore,
designers are compelled to
also employ
soft Intellectual Properties (IPs) for e.g., multiplication and division functions, proposed by major FPGA vendors such as Xilinx and Intel \cite{XilinxMultIpCore, XilinxDivIpCore}. In fact, exploitation of soft IPs instead of DSPs, for low bit-width operations, has also been suggested by academia and industry \cite{XilinxDSPCore, 8978722}.
Nonetheless, the long latency and high resource footprint of LUT-based IPs still needs to be decreased to facilitate the deployment of off-the-shelf FPGAs in aerial platforms and wearable gadgets.
On the other hand, to boost the throughput in despite of stagnant clock speed, \textit{pipelined} IPs have been proposed as a promising solution, albeit leaving the
the quest for reducing the resource-gap unaddressed.

\begin{figure}[t]
 \centering
  \includegraphics[width=0.38\textwidth]{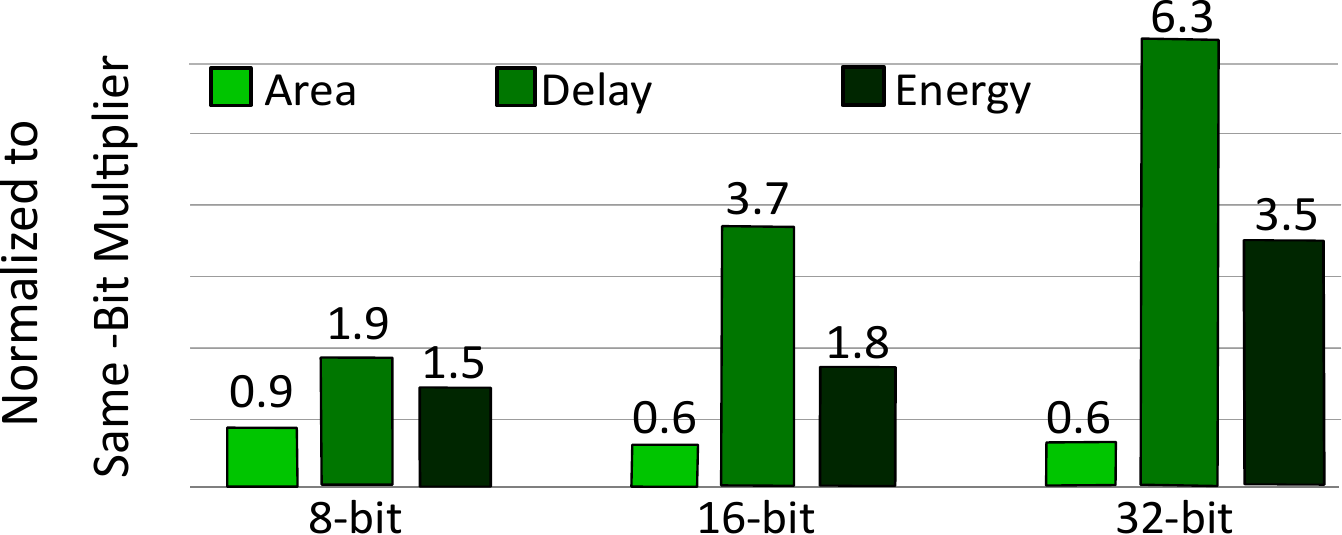}
     \vspace{-0cm}
 \caption{Comparing area, delay, and energy of 8-, 16-, and 32-bit multiplication and division functions, implemented in Virtex-7 FPGA through Look-up Table (LUT) based accurate IPs.
 }\label{fig:ALU_breakdown}
  \vspace{-0.5cm}
\end{figure} 
\vspace{-0cm}

To minimize the resource footprint in aforementioned error-resilient programs, many approximation techniques have been emerged,
however, three main challenges are attributed to them. First, approximation approaches customized in ASIC platforms have not yielded comparable performance gains when directly synthesized and ported to FPGAs, due to their different architectural specifications
\cite{10.1145/3195970.3195996}. 
Second, approximation is often applied on a single kernel of a multi-kernel application, e.g., replacing multiplication (or division) with imprecise versions in the DCT (quantization) stage of JPEG compression. Their \textit{performance gain} is also often evaluated and reported for that single kernel and not in the end-to-end implementation of the application.
Third, while most efforts are concentrated on multiplication, studies like \cite{10.1145/3386263.3406907, 8807077} and our analysis shown in Fig. \ref{fig:ALU_breakdown}, consider the longer latency and higher energy of division compared to the multiplication. In fact, division is often the speed-bottleneck arithmetic operation in soft processors and can constrain the speed of application. Therefore, approximation of division has also become pronounced as, although less frequent, this operation is unavoidable in image processing and vision applications (Harris corner detection and K-means in unsupervised clustering) as well as bio-signal processing (heartbeat detection). In this context, a pipelined/low-latency divider is of high desire. Nevertheless, among the imprecise dividers in the literature
only one is targeted for LUT-based architectures \cite{10.1145/3386263.3406907} and none is specialized toward a pipelined design. In fact, pipelining has been mostly applied, in coarse-grain granularity, in a processor datapath while the fine-grained approach has been overlooked, especially in approximate designs.

To cope with the first challenge, we have designed \textit{LeAp} \cite{9045171} as the first logarithmic multiplier specialized for FPGAs. The motivation behind devising LeAp is based on translation of multiplication to addition in the logarithmic presentation, through 
Mitchell's algorithm \cite{5219391} \small{($P = A \times B \xrightarrow[\text{Log}]{\text{Approx.}} \widetilde{Log_P} = \widetilde{Log_A} + \widetilde{Log_B} \xrightarrow[\text{Anti-Log}]{\text{Approx.}} \tilde{P} = 2 ^{\widetilde{Log_P}}$)}. Transforming the 2D array structure of multiplication to 1D addition
significantly reduces the design complexity. Specifically, it reduces the long latency and energy of an accurate divider, close to those of an accurate multiplier of the same-size.
This translation is also suited for FPGAs and enables substantial gains as they already encompass
fast carry chains
to expedite addition. Implementation of Mitchell's algorithm consists of three steps: \normalsize approximate the log of inputs (by finding the position of the leading one), addition of two logs, and finally anti-log (a shift operation). 
LeAp has taken major leaps toward speeding-up Leading One Detection (LOD) and error-reduction steps: in the former, the LOD is accelerated by probing the 4-bit segment in parallel, and then finding the leading one in the most significant segment. For the latter, we have augmented the original Mitchell's algorithm with three novel error-reduction schemes (independent from multiplier-size). The proposed error-reduction schemes allow the \textit{addition} of the error coefficient \textit{concurrently} with the fractional addition step and within the same resource footprint, used for implementing the baseline Mitchell's designs \cite{9211770}.
Furthermore, LeAp \cite{9045171} is customized in a way that effectively utilizes FPGA primitives, based upon the fact that LUTs and carry chain (having fixed latency),
can also be configured to perform a ternary addition (frac$_1$+frac$_2$+error coefficient). 
This is in contrast to previous approaches (MBM \cite{8493590} and INZeD \cite{8807077}), for which an additional circuitry was necessary for the addition of error-reduction terms to the original Mitchell's circuits.


To surmount the rest of foregoing challenges in FPGAs, we further expand LeAp multiplier \cite{9045171} and propose \textit{\pname}, which sets out to enable the first architectures for approximate \textit{fine-grained pipelined} multiplier and divider. Our novel contributions in \pname~architectures are highlighted from three perspectives: first, following the light-weight error-reduction scheme of LeAp, we propose a logarithmic divider. Second, we further boost the accuracy of our designs by proposing various configurations for both multiplication and division that also surpass the SoA counterparts in terms of accuracy and/or performance metrics.
Finally, we tailor our non-pipelined architectures for fine-grain pipelining and propose
various versions of approximate multiplier and divider, having different number of stages. Such a spectrum of configurations also enables diverse power-throughput trade-offs, suited for different operating frequency levels within FPGAs. It is worth underlining that featuring fine-grained pipelining at an intra-unit granularity could enable operating at a higher frequency-level versus coarser-granularity (i.e., inter-unit), especially in applications including
division operation.
Note that although our circuits are specialized for FPGAs, our fine-grain pipelining approach is also applicable to ASIC-based architectures.
In short, we make the following key technical research contributions in this journal version:

\begin{itemize}[leftmargin=*]

\item \textbf{Near-zero biased multiplier\,\&\,divider with extendable accuracy}. We further increase accuracy of LeAp multiplier and propose three versions of divider. The accuracy of error-reduction schemes are further increased so that the minimally-biased designs confine the average of absolute relative error (ARE) from $>$3.8\% to $<$0.6\%, through a limited number of coefficients.

\item \textbf{First approximate pipelined multiplier and divider}. We design 2- 3-, and 4-stage \textit{pipeline} architectures that achieve up to 8.1$\times$ higher throughput and 6.8$\times$ higher throughput/Watt over accurate IP-based counterparts from Xilinx Vivado.

\item \textbf{End-to-end performance-QoR evaluation on three application domains}. The efficacy of the proposed multiplier and divider is evaluated in the end-to-end implementation of three applications: Pan-Tompkins heartbeat QRS detection, JPEG compression, and Harris Corner Detection (HCD). 

\item \textbf{Open-source model}. The implementations of \pname~multipliers, dividers, and FPGA-customized programs will be open-sourced at \href{https://cfaed.tu-dresden.de/pd-downloads}{\textcolor{blue}{\textit{https://cfaed.tu-dresden.de/pd-downloads}}}, to fuel further research in reconfigurable and approximate computing communities.
\end{itemize}



The rest of this article is organized as follows: Section \ref{sec:related} presents a brief survey on the imprecise multipliers and dividers and distinguish the contribution of this work. Section \ref{sec:background} summarizes a background on Mitchell's Mul/Div algorithms
We
elaborate upon the proposed architectures and pipelining approach in Section \ref{sec:proposed}, respectively. Experimental setup, circuit- and application-level results (on three multi-kernel applications) are detailed in Section \ref{sec:results}. Finally, Section \ref{sec:conclusion} draws the conclusion
with an outlook to interesting future tracks.

\section{Related Work} \label{sec:related}
Although substantial effort has been dedicated to
ASIC-based inexact multipliers and dividers (a quantitative evaluation of them can be found in \cite{9165786}), FPGA-specialized counterparts have also recently gained traction. The approaches in both landscapes are discussed herein and their compendium is highlighted in Table \ref{table:related}. 

\textbf{Partial product (PP) approximation}: a wide class of works targeting imprecise multipliers have mostly applied approximation on: 1) PP generation by simplification of truth table in e.g., 2$\times$2 and 4$\times$4 multipliers and use them in a modular design \cite{7827657}. 
2) Reduction/accumulation of PP rows into two, using 3:2 and 4:2 imprecise compressors \cite{9181066, 9086885, 9026985, 8383694, 8978722, 7817891, 7817900} in Dadda- or array-based multipliers. or 3) Adding the reduced PP rows by e.g., splitting the carry propagation path \cite{10.1145/3195970.3195996, 9045546}.
The main drawback of these approaches is weak-scalability to larger input-width,
as error may drastically increase, when accumulated in a hierarchical design approach. Furthermore, compressor-based designs are posed with high average relative error (ARE), or render moderate performance gain when applied on few least significant PP columns.

\begin{table*}[!t]
\centering
\caption{{Summary of state-of-the-art approximation approaches for ASIC- and FPGA-based multipliers and dividers \cite{10.1145/3486616}
}}
\vspace{-0.1 cm}
\label{table:related}
\begin{threeparttable}
\def\arraystretch{1.2}
\resizebox{\textwidth}{!}{
\begin{tabular}{|c|c|c|c|c|c|c|c|}
\Xhline{5\arrayrulewidth}
\textbf{Approach}                                                                                                            & \textbf{Mul/Div}          & \textbf{Pipelined}   & \textbf{\begin{tabular}[c]{@{}c@{}}ARE$^\textbf{1}$\\ up to (\%)\end{tabular}} & \textbf{Description}                                                                                          & \textbf{\footnotesize{Platform}}     & \begin{tabular}[c]{@{}c@{}} \textbf{Reported} \\ \textbf{Mul/Div Gain$^\textbf{2}$} \end{tabular}                        & \textbf{End-to-end$^\textbf{4}$} \\ \Xhline{5\arrayrulewidth}
\multirow{8}{*}{\textbf{\begin{tabular}[c]{@{}c@{}}Partial\\ Product\\ Generation/\\ Addition/\\ Accumulation\end{tabular}}} & \multirow{7}{*}{\cmark/\xmark} & \multirow{7}{*}{\xmark} & 7.6                                                               & Inexact 4:2 compressor in Dadda Mul \cite{9086885, 9026985, 8383694, 7817900, 7817891}  & \multirow{4}{*}{ASIC} & \{A, D, P\} +                     & In \cite{7817891}                                               \\ \cline{4-5} \cline{7-8} 
                                                                                                                             &                           &                     & 1.7                                                               & Asymmetrically utilize inexact compressor in 3/4 of LSB columns \cite{9181066}                                 &                       & \{A, P\} ++, D +                  & \multirow{6}{*}{\xmark}                                                   \\ \cline{4-5} \cline{7-7}
                                                                                                                             &                           &                     & 8.4                                                               & Simplified Karnaugh map of 4:2 compressor in Booth multiplier \cite{8353383}                                      &                       & A +, E ++                            &                                                                            \\ \cline{4-5} \cline{7-7}
                                                                                                                             &                           &                     & Config.                                                      & Library of larger multiplier and adders using 2x2 instances \cite{7926993, 7827657}                            &                       & \{A, D, P\} ++                    &                                                                            \\ \Xcline{4-7}{1.3pt}
                                                                                                                             &                           &                     & 0.3                                                               & Cutting the carry propagation path in 4-, 8-bit array multiplier \cite{9045546, 10.1145/3195970.3195996}                                          & \multirow{3}{*}{FPGA} & \{A, D, P\} +                     &                                                                            \\ \cline{4-5} \cline{7-7}
                                                                                                                             &                           &                     & 8.5                                                               & Truth table simplification 3:2/4:2 compressor for Dadda multiplication \cite{8978722}                      &                       & \{A, P\} +                            &                                                                            \\ \cline{4-5} \cline{7-7}
                                                                                                                             &                           &                     & Config.                                             & Library of 4x4 and 8x8 with approximate partial products \cite{8465845}                                       &                       & \{A, P, D\} +                           &                                     \\\Xhline{5\arrayrulewidth}
\multirow{3}{*}{\textbf{\begin{tabular}[c]{@{}c@{}}Truncated\\ Mul/Div\end{tabular}}}                                        & \cmark/ \xmark                 & \multirow{2}{*}{\xmark} & 10.9                                                              & Leading one based: with error compensation \cite{9105108}, with rounding \cite{8626488, 7517375}   & \multirow{3}{*}{ASIC} & \{A, P\} ++                           & \multirow{3}{*}{\xmark}                                                   \\ \cline{2-2} \cline{4-5} \cline{7-7}
                                                                                                                             & \xmark/ \cmark                 &                     & 6.7                                                               & Leading-one position based 2k+2/k+1 Div plus error reduction circuit \cite{8713893, 8342233}                   &                       & \{A, D\} +, E ++                 &                                                                            \\ \cline{2-5} \cline{7-7}
                                                                                                                                                                                                                                                          & \cmark/ \xmark                 &                \xmark     & 1.2-4.7                                                               & Variable-precision multiplier based on 8-bit truncated instances \cite{8713893, 8342233}                   &                       & \{A, E\} +                 &                                                                            \\ \Xhline{5\arrayrulewidth}
\multirow{4}{*}{\textbf{\begin{tabular}[c]{@{}c@{}}Multiplicative\\ Dividers\end{tabular}}}                                  & \multirow{4}{*}{\xmark/\cmark} & \multirow{3}{*}{\xmark} & 2.9                                                               & Piecewise linear approximation and rounding of reciprocal of divisor \cite{10.1145/3218603.3218650}            & \multirow{4}{*}{ASIC} & \{D, E\} ++                         & \multirow{3}{*}{\xmark}                                                   \\ \cline{4-5} \cline{7-7}
                                                                                                                             &                           &                     & 6.4                                                               & Approximating reciprocal by bit manipulation \cite{7927254}, with truncation \cite{7927254}                     &                       & \{A, D, E\} +++                  &                                                                            \\ \cline{4-5} \cline{7-7}
                                                                                                                             &                           &                     & 16.3                                                              & Approximating reciprocal using a table indexed by upper bits of divisor \cite{7459545}                         &                       & \{A, D\} +, E ++                 &                                                                            \\ \cline{3-5} \cline{7-8} 
                                                                                                                             &                           &         \cmark       & 4.9                                                               & Incremental approximation of reciprocal of divisor using Taylor series \cite{8766885}
                                                                                                                            &                       & \{D, E\} +++                        & \xmark                                                               \\ \Xhline{5\arrayrulewidth}
\multirow{8}{*}{\textbf{\begin{tabular}[c]{@{}c@{}}Mitchell’s\\ Multiplication\\ and Division\\ Algorithms \\ \cite{5219391} \end{tabular}}}    & \multirow{4}{*}{ \cmark/ \xmark   }              & \multirow{5}{*}{ \xmark  }               & 2.9                                                               & Enhance Log accuracy: round rather truncation in piecewise approximation \cite{8714868}                 & \multirow{5}{*}{ASIC  }                & \{A, P\} ++, D +                  & \multirow{5}{*}{\xmark  }                                                                 \\ \cline{4-5} \cline{7-7} 
                                                                                                                             &                           &                     & $>$ 3.9                                                  & Use different approximate adders in Mitchell's multiplier \cite{8280549}                                       &                       & \{A, P\} +++                          &                                                                            \\ \cline{4-5} \cline{7-7}
                                                                                                                             &                           &                   &                2.7                                                   & Improving accuracy of Mitchell's Mul with adding one error-correction  \cite{8493590}                          &                       & \{A, P\} ++                           &                                                                            \\ \cline{4-5} \cline{7-7}
                                                                                                                             &                           &                     & 2.7                                                               & Adding up to 256 error-coefficient to Mitchell's muliplier \cite{9116315}                                      &                       & \{A, P\} +                            &                                                                            \\ \cline{2-2} \cline{4-5} \cline{7-7} 
                                                                                                                             & \xmark/ \cmark                 &                     &             3.0                                                     & Add one error-correction (with a similar approach to \cite{8493590}) \cite{8807077}                             &                       & A +, \{D, E\} ++                 &                                                                            \\ \Xcline{2-8}{1.3pt} 
                                                                                                                         & \cmark/ \xmark                 &       \xmark              &                                        1-1.6                           & Adding one to five error-coefficient to Mitchell's multiplier \cite{9045171}                                    &   
                                                                                                                       FPGA
                                                                                                                        & \{A, E\} ++, T +            &  
                                                                                                                        \xmark  
                                                                                                                        \\ \cline{2-5} \cline{6-8} 
                                                                                                                                                                                                                                                      & \cmark/ \cmark                & \xmark                &          0.8                                                        & Adding 64 error-coefficients to Mitchell's multiplier and divider \cite{10.1145/3386263.3406907, 10.1145/3486616, 9401461}               &  FPGA/ASIC  & \{A, E\} ++, T +++             &      In \cite{10.1145/3486616, 9401461}                                                            \\ \cline{2-8} 

\rowcolor{lightgreen}                                                                         \cellcolor{white}         &  \textbf{\cmark/\cmark}        &   \cmark    & \textbf{0.6-1}                                                     & \textbf{Different pipelined designs for multiplier and divider with tunable accuracy}                           &       \textbf{FPGA}              & \textbf{\{A, E\} +++, T +++} & \textbf{\cmark}                                                         \\ \Xhline{5\arrayrulewidth}
\end{tabular}
}
\begin{tablenotes}
\item $^1$\,Average of Absolute Relative Error (a.k.a MRED),  \blank{0.3cm}  $^2$\,\underline{A}rea/\underline{D}elay/\underline{E}nergy/\underline{P}ower/\underline{T}hroughput
\blank{0.3cm}
$^4$\,End-to-end performance gain
\end{tablenotes}
\end{threeparttable}
\vspace{-0.2 cm}
\end{table*}

\textbf{Resizing to narrower-width multiplier/divider}: another category of works utilize a smaller instance of the arithmetic unit, based on a dynamic selection of most significant bits, starting from the position of the leading one (\cite{8626488, 9105108, 7517375, 7372600} for multiplier and \cite{8713893, 8342233, 7544348} for divider). Although offering high resource improvements, these truncation techniques suffer from error cases up to 100\%. In addition, the latency of an accurate smaller-divider could be still multiple times of a same-sized multiplier.

\textbf{Division with inexact subtractor}: this branch of studies have truncated or replaced accurate subtractors with imprecise counterparts, in 2N-by-N non-restoring and restoring array dividers \cite{8902027, 8702363, 8322284, 7303928}. The main difference between the two is in the remainder output, non-restoring version does not correct the 
remainder when subtraction has a negative result. Thus, a correction circuit is added to its hardware implementation.
It has been shown that an approximate restoring version
dissipates less power than its non-restoring counterpart,
while introducing slightly larger accuracy degradation \cite{7303928}.
Such cells are also exploited in combinational implementation of higher radix SRT dividers. Generally approximate array dividers offer high accuracy, however, the resource savings achieved by them are small due to their array structure \cite{8713893, 8342233}. Furthermore, their latency remains multiple times of a same-sized multiplier.

\textbf{Multiplicative dividers}: in another class of dividers, the reciprocal of the divisor is calculated first and subsequently multiplied with the dividend. Approximation is applied on the reciprocal of the divisor (using Taylor series \cite{10.1145/3287624.3287668} or linear piecewise approximation \cite{10.1145/3218603.3218650}). Encoding the reciprocal to lie in the range of (0.5, 1] is performed usually using a table indexed by the upper bits of the divisor (similar to \cite{7459545}), or a series of bit manipulations, e.g., inverting all bits and appending '1' at MSB (like \cite{7927254}). In some studies operands are also truncated/rounded to lie in some specific bit-width range. Overall, these approaches impose significant resource cost in FPGAs, due to requiring both reciprocal and multiplier IPs, separately. Moreover, when truncating divisor goes beyond relatively few bits, the accuracy of the reciprocated divisor degrades significantly \cite{10.1145/3218603.3218650}.

\textbf{FPGA-customized multipliers}: it has been shown that ASIC-based approximation approaches have not yielded comparable performance gains when directly synthesized and ported to FPGAs \cite {10.1145/3195970.3195996, 8465845} (primarily, due to the underlying architectural differences).
To cope with this challenge, recent works narrowed their focus toward
specializing such techniques taking into account the LUT-based structure of FPGAs. For example, 
modification of LUTs which calculate LSBs in 4$\times$4 multiplications have been customized separately, for array based- \cite {9045546, 10.1145/3195970.3195996, 8465845}, and Booth-based \cite{9072581} architectures. Furthermore, an approximate compressor has been proposed in \cite{8978722}, geared toward an LUT-oriented implementation. Despite the specialized customization in these schemes for reconfigurable platforms, their resource savings have not been significant compared to logarithmic designs (discussed in the following). Moreover, Booth-based designs suffer from an increased critical path length \cite{9072581}. This has highlighted the need for exploring other avenues for FPGAs which can achieve higher performance gain with an acceptable accuracy.

\textbf{Logarithmic multiplier and dividers}: In a more recent trend, some works have adopted Mitchell's approximate algorithms in lieu of array-based designs. Mitchell's algorithms translate multiplication (division) into log and addition (subtraction) in logarithmic representation.
Such logarithmic transformations generally render higher resource gains compared to other approximation approaches.
Interestingly, the latency of a division is also reduced, comparable to a same-sized multiplier, via this algorithm.
However, these improvements come with the cost of relatively high error (ARE of 3.8\% for multiplication and 4.1\% for division). Therefore, various schemes have been presented to reduce the error in the original Mitchell's algorithms. 
The authors of MBM \cite{8493590} have proposed a single error-reduction term for multiplication, and they employed a similar scheme for division in INZeD \cite{8807077}. However, a single error-reduction term weakly fits all input combinations and eventuates in many output overflow cases, when adding the error-reduction term. In addition, their approaches are optimized for ASIC platforms. To alleviate error and also targeting FPGAs, we have recently proposed two FPGA-specialized designs, LeAp \cite{9045171} and SIMDive \cite{10.1145/3386263.3406907} (a similar error-reduction approach to SIMDive has also been adopted in the REALM multiplier \cite{9116315}). The approaches in these works considers $F$ MSBs of fractional parts for each operand. Therefore, the possible combinations for the possible pairs makes a squarish region which is then partitioned to $2^F\times2^F$ sub-regions, each assigned with a unique error-coefficient.
Although the error-reduction approaches of these works are similar in spirit, each has been designed for either ASIC or FPGA and targeted for different design goal. While REALM \cite{9116315} is an ASIC multiplier, SIMDive \cite{10.1145/3386263.3406907} implements an FPGA-specific hybrid Mul/Div, aimed at applications with data parallelism opportunities that can benefit from Single Instruction, Multiple Data (SIMD) architectures. Moreover, as will be shown later, \pname~enables higher accuracy than SIMDive, even with smaller number of coefficients/LUTs.


\textbf{Pipelined architectures}: \textit{Fine-grain} pipelining has recently gained attention. In this context, authors in \cite{8715045, 8877443} have applied pipelining on PP generation and accumulation in the array-based \textit{accurate} multipliers.
Compared to accurate-, limited attention has been dedicated to approximate-designs. SoA divider SAADI-EC \cite{8766885} and a modified version of the DRUM multiplier \cite{9547961} have been the only works that adopted intra-unit pipelining. Despite the possibility of increasing throughput, these work lack a well-established pipelining strategy as no analysis has been performed to uniformly divide the critical path over pipeline stages, resulting in poor performance.
To the best of our knowledge, no other work has proposed approximate pipelined multiplier or divider architectures, which is highly desired for processing of cutting-edge applications. In fact, the achieved substantial boost in throughput is pivotal in e.g., image processing and health monitoring applications.








\section{Preliminaries and Background}  \label{sec:background}

\textit{Mitchell's Multiplication and Division Algorithms}: as shown in Eq. \ref{equation0}, Mitchell's algorithms perform imprecise multiplication and division in the logarithmic representation of numbers. Consider the binary representation for $N$-bit unsigned input $A$, which can be written as Eq. \ref{equation1_1}, where $k$ indicate
the position of the leading one. The rest of the bits (starting from position $k-1$ to $0$) are considered as the fractional part and fall in the range of $0\leq x<1$. \vspace{-0.1cm}

\begin{equation}
\vspace{-0.1cm}
\label{equation0}
\begin{cases}P = A \times B \xrightarrow[\text{Log}]{\text{Approx.}} \widetilde{Log_P} = \widetilde{Log_A} + \widetilde{Log_B} \xrightarrow[\text{Anti-Log}]{\text{Approx.}} \tilde{P} = 2 ^{\widetilde{Log_P}} \\\\D = A \div B \xrightarrow[\text{Log}]{\text{Approx.}} \widetilde{Log_D} = \widetilde{Log_A} - \widetilde{Log_B} \xrightarrow[\text{Anti-Log}]{\text{Approx.}} \tilde{D} = 2 ^{\widetilde{Log_D}} \end{cases}  
\end{equation}


\setlength{\abovedisplayskip}{-2pt}

\footnotesize
\begin{equation}
\label{equation1_1}
\!\!\!\!\!\!\!\!\!\!A=2^k\!+\!\sum_{i=0}^{k-1}2^ib_i=2^k(1+x)\xrightarrow{e.g.} 58=2^5(1+0.11010)_2,18=2^4(1+0.001)_2\!\!\!
\end{equation}
\normalsize

In linear mathematics,\,$log_2(1 + x)$\,is approximated to $x$ for this range of $0\leq x<1$; thus, the approximate log value of input $A$ is: 

\footnotesize
\begin{equation}
\label{equation2}
Log_2(A) \simeq k+x 
\rightarrow Log_2(58)\simeq(101.11010)_2, Log_2(18)\simeq(100.001)_2
\end{equation}
\vspace{-0.4cm}
\normalsize

After applying the same step on the second input to get its approximate log, the summation (subtraction) of two parts is obtained in Eq. \ref{equation3_1} (Eq. \ref{equation3_2}).

\footnotesize
\begin{equation}
\label{equation3_1}
\widetilde{Log_2}(\tilde{P})=(k_1 + k_2) + (x_1+x_2)
\small\rightarrow K_{s}=(1001)_2, X_{s}=(0.1111)_2 
\end{equation}

\vspace{-0.2cm}

\begin{equation}
\label{equation3_2}
\widetilde{Log_2}(\tilde{D})=(k_1-k_2)+ (x_1-x_2)
\rightarrow  K_{s}=(1)_2, X_{s}=(0.1011)_2
\end{equation}
\normalsize
Finally, by applying the anti-log (which mathematically is a shift operation), binary representation of the approximate product (quotient) are derived by Eq. \ref{equation4_1} (Eq. \ref{equation4_2}): 


\footnotesize
\begin{equation}
\label{equation4_1}
\!\!\!\!\!\!\!\!\!\!\widetilde{P}\!\!=\!\!\left\{\begin{matrix}
 2^{k_1+k_2}(1+x_1+x_2), & \!\!\!\!\!\!\!\!\!\! x_1+x_2<1 \\  
 2^{k_1+k_2+1} (x_1+x_2), \!\!\!\!\!\!\!\!\!\! & x_1+x_2\geq1
\end{matrix}\right.
\rightarrow \widetilde{P} 
= 992, P_{acc}= 1044
\end{equation}


\begin{equation}
\label{equation4_2}
\!\!\!\!\!\!\!\!\!\!\widetilde{D}\!\!=\!\!\left\{\begin{matrix}
 2^{k_1-k_2-1}(2+x_1-x_2), & \!\!\!\! x_1-x_2<0 \\  
 2^{k_1-k_2} (1+x_1-x_2),  & \!\!\!\! x_1-x_2\geq0
\end{matrix}\right.
\rightarrow  \widetilde{D} = (11)_2 = 3,    D_{acc}= 3    
\end{equation}

\normalsize

\vspace{0.2cm} 
\section{Proposed Pipelined and Non-Pipelined Multiplier and Divider Architectures} 

\label{sec:proposed}
In this section we initially present the proposed error-reduction schemes, and afterwards elaborate on the structure of \pname~multiplier and divider architectures (non-pipelined and pipelined).

\subsection{Proposed\;light$\mbox{-}$weight\,\&\,minimally$\mbox{-}$biased\;error$\mbox{-}$reduction\;schemes}
Mitchell's error for 8-bit multiplication and division are formulated in the Equation \ref{equation5_1} and \ref{equation5_2}.
Inspecting the behavior of the error has provided the following insights, some of which are also noted and endorsed by SoAs \cite{9116315, 10.1145/3386263.3406907, 8493590, 8807077}:

\footnotesize
\begin{equation}
\label{equation5_1}
E_P=P-\widetilde{P}=\left\{\begin{matrix}
 2^{k_1+k_2}(x_1x_2), & x_1+x_2<1\\ 
 2^{k_1+k_2} (1-x_1-x_2+x_1x_2),  & x_1+x_2\geq1
\end{matrix}\right.
\end{equation}
  \vspace{-0.1cm}

\begin{equation}
\label{equation5_2}
E_D=D-\tilde{D}=\left\{\begin{matrix}
 2^{k_1-k_2}\, \frac{(x_1(x_2-1)+x_2-(x_2)^2)}{2(1+x_2)}, & x_1-x_2<0\\ 
 2^{k_1-k_2} \, \frac{(x_1x_2-(x_2)^2)}{1+x_2},  & x_1-x_2\geq0
\end{matrix}\right.
\end{equation}

\begin{figure*}[t]
 \centering
  \includegraphics[width=1.01\textwidth]{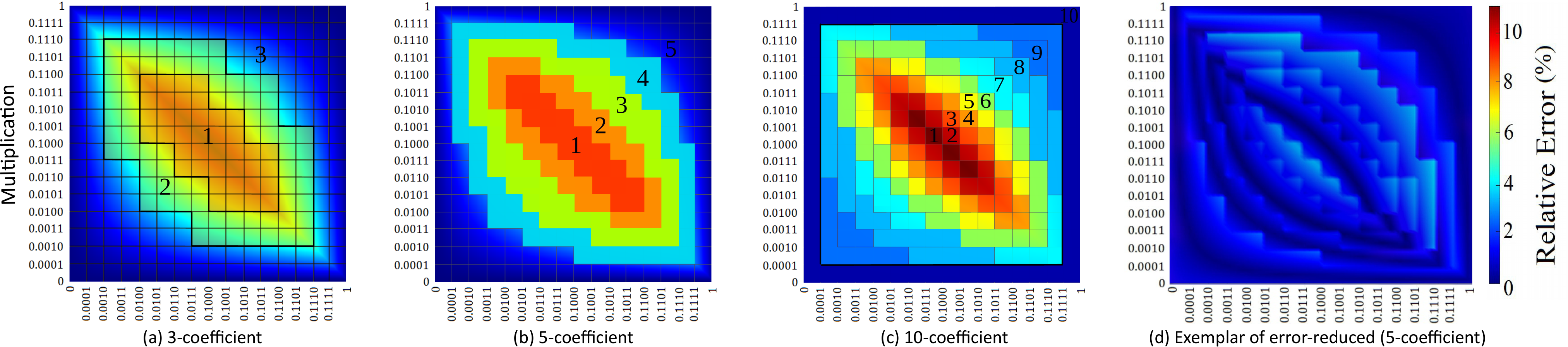}
      \vspace{0.2cm}
    \includegraphics[width=1.01\textwidth]{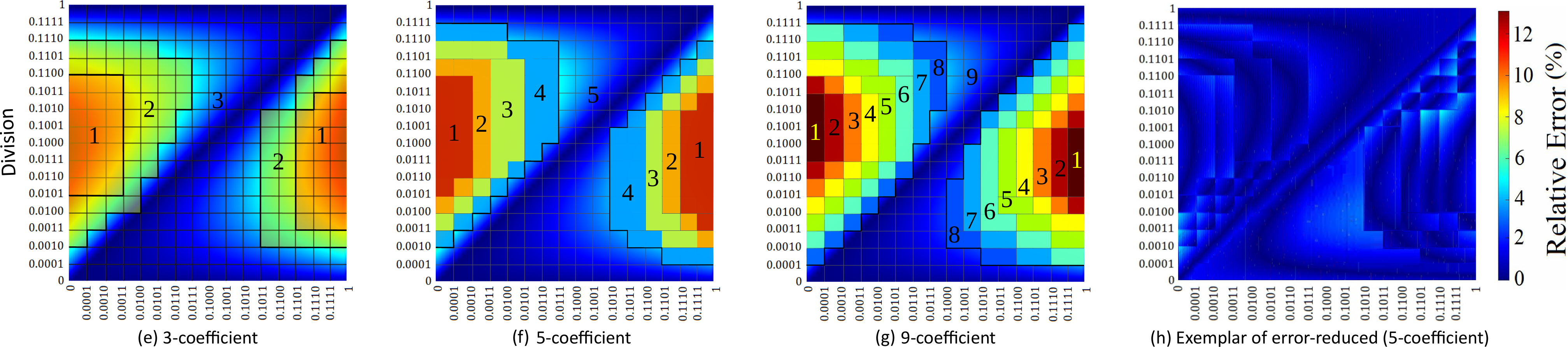}
     \vspace{-0.4cm}
 \caption{Proposed error reduction schemes of \pname~for multiplication and division, based on 4 MSBs of fractional parts.
}\label{fig:coefficients}
 \vspace{0.1cm}
\end{figure*}

\begin{itemize}[leftmargin=*]
\item
The equations demonstrate the different error magnitude in each power-of-two interval. Therefore, merely adding a single correction term to the output (as proposed in INZeD \cite{8807077} and MBM \cite{8493590}) is not efficient and results in many output overflow cases \cite{9116315, 10.1145/3386263.3406907}.
 
\item
The equations also reflect the proportional replication of error in each power-of-two. This behavior is repeated for each size of multiplication or division (irrespective of $k_1$ and $k_2$). This repetition allows applying the same error-reduction approach for all multiplier or divider sizes. In fact, a set of coefficients can be added to fractional parts, before that their summation become scaled \cite{8493590, 8807077}).

\item As can be seen in Fig. \ref{fig:coefficients} (a), and (e), some of nearby sub-regions have very small error (less than 2\%), while the changes of error occurs is more pronounced in other sub-regions. This motivates grouping the regions having similar error as well as assigning more coefficients to the sub-regions having higher error (to efficiently reduce the error to a certain bound).

\end{itemize}

\begin{table*}[t]
\large
\centering
\caption{{Binary representation of error-reduction coefficients in 16-bit multiplier \& divider (3/4 MSBs are zero for Mul/Div and excluded)}
}
\vspace{-0.1 cm}
\label{table:coefficients}
\def\arraystretch{1.1}
\resizebox{\textwidth}{!}{
\begin{tabular}{|c|c|c|c?{1mm}c|c|c|c|}
\Xhline{4\arrayrulewidth}
\multicolumn{4}{|c?{1mm}}{\textbf{Multiplier}}                            & \multicolumn{4}{c|}{\textbf{Divider}}                                \\ \Xhline{5\arrayrulewidth}
\textbf{3-coefficient} & \multicolumn{1}{c|}{\textbf{5-coefficient}} & \multicolumn{2}{c?{1mm}}{\textbf{10-coefficient}} & \textbf{3-coefficient} & \textbf{5-coefficient} & \multicolumn{2}{c|}{\textbf{9-coefficient}} \\ \Xhline{3\arrayrulewidth}
1) 100000100111     & 1) 1001111111111                         & 1) 1001111000110      & 6) 0101110011111     & 1) 1000011111111    & 1) 1001111000100    & 1) 1001110001111      & 6) 0110010100101     \\ \hline
2) 010011101100     & 2) 1000011011101                         & 2) 1000110110001      & 7) 0100101000011     & 2) 0100010111111    & 2) 1000001000111    & 2) 1000110111100      & 7) 0101000101011     \\ \hline
3) 000100101001     & 3) 0110010001010                         & 3) 0111111000100      & 8) 0100001011101     & 3) 0001011111111    & 3) 0110110001101    & 3) 1000000010100      & 8) 0100111101000     \\ \hline
                 & 4) 0011110010111                         & 4) 0111000110101      & 9) 0011110000011     &                  & 4) 0101010100111    & 4) 0111001100010      & 9) 0100001101100     \\ \hline
                 & 5) 0000111110000                         & 5) 0110010100011      & 10) 0010101111111     &                  & 5) 0011011100100    & 5) 0110100001101      &     \\ \hline

\end{tabular}
}
\vspace{-0.2 cm}
\end{table*}

\begin{figure}[t]
 \centering
  \includegraphics[width=0.49\textwidth]{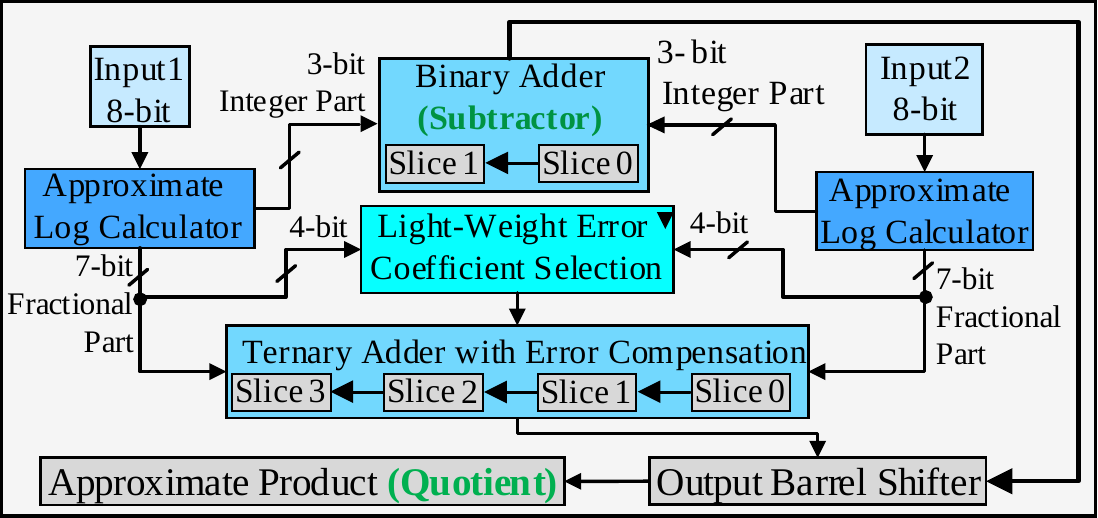}
     \vspace{-.5cm}
 \caption{Overall structure of proposed \pname~multiplier and divider
}\label{fig:proposed_structure}
  \vspace{-0.5cm}
\end{figure}

It is also worth noting that an efficient partitioning should result in a small number of coefficients to minimize the overhead of error-reduction. Partitioning approaches such as REALM \cite{9116315} and SIMDive \cite{10.1145/3386263.3406907} results in superfluous number of error-coefficients when targeting high accuracy. Note, considering even four MSBs of fractional parts results in $2^4\times 2^4=256$ coefficients and limits the scalability of such schemes (especially for FPGAs, as will be discussed later).
It fact, the resource-efficiency of the error-reduction strategy proposed in SIMDive depends on the number of LUT inputs and is not easily extendable when when targeting higher accuracy. Going into the details, to generate 256 coefficient through SIMDive approach, four MSBs from each fractional part should be considered, which requires the direct utilization of 8-LUTs for each bit of error-reduction term. Otherwise the usage of a 256-to-1 Mux (each input is 8- or 16-bit) is inevitable. Implementing this mux or 8-LUT out of 6-LUTs increases the error-reduction overhead of SIMDive to nearly 4$\times$ of its original idea (well-suited for considering 3 or less MSBs of fractional part). This overhead is not negligible, especially when targeting a small-sized multiplier or divider. Building upon this discussion, we propose \pname~to \textit{enable reaching higher accuracy with even smaller overhead.}

\textbf{Proposed light-weight error-reduction scheme}: to cope with the overflow problem in INZeD and MBM, and the over-provisioned number of coefficients (i.e., 256) in REALM, the insights obtained from aforementioned observations incentivize reducing the error-reduction terms by efficiently clustering them in fewer groups. In \pname, we divide the squarish region between each power-of-two pair to \textit{different, and fewer}, regions than REALM/SIMDive. In our partitioning we consider the following factors: \\
1) Consider four rather than three MSBs of fractional parts, to increase the accuracy.\\
2) Minimize the number of sub-regions while considering four MSBs, to constrain resource cost for selection of the coefficient.\\
3) Minimize $error\;distribution\times error\;magnitude$ in each region (can be estimated to the integral of error-magnitude for that region).\\
We have proposed three error-reduction schemes in Fig. \ref{fig:coefficients}, based on the in-depth analysis of error and the resource-usage reports from Vivado. To minimize the average of absolute relative error (ARE), various partitionings have been investigated.
The goal of partitioning is to keep the error of grouped sub-regions nearly uniform and below pre-defined thresholds, e.g., $\sim$4\%, 3\%, and 2.5\% for 3-, 5-, and 10-coefficient multiplier schemes, respectively.
Further, the error-reduction coefficient for each group of sub-intervals, are derived by following the mathematical approach detailed in \cite{9116315}.
The binary representation of the proposed coefficients for multiplier and divider are also shown in Table \ref{table:coefficients}.
From the implementation point of view, partitioning is implemented via small-sized multiplexers (coded via casex statement in HDL). In such implementation, the LUT usage depends on \textit{the number of error-coefficients as the Mux-inputs} and \textit{the complexity of conditional statements in the casex statement}. In order to minimize the former (number of Mux-inputs) we have proposed three schemes, having most 10 coefficients. In order to minimize the latter (complexity of conditional statements for selecting the coefficients), we only consider comparing 4 MSBs of fractional parts for the partitioning. Moreover, in this partitioning, neighbouring sub-regions having the same coefficient are packed to reduce the complexity for each conditional statement.
From the hardware-usage perspective, each 6-LUT functions as a 4-MUX (in which 4 inputs and the 2 select lines are the LUT inputs). Further, as also denoted by \cite{Multiplexer_FPGA}, a 16:1 multiplexer requires a single FPGA-slice having four 6-LUTs. Based upon this, we have also chosen a a squarish partitioning through a MUX-based approach, to minimize the overhead for the proposed error-reduction strategy. This is while, checking the rhombus borders (see the original error shape in Fig. \ref{fig:coefficients}) is very costly in hardware and could nullify the LUT-saving, gained from approximating approach.
It should be noted that the proposed partitioning is also scalable, as the resource cost for choosing one of few coefficients does not grow exponentially, contrary to the REALM/SIMDive approaches. In fact, the proposed partitioning surpasses the SoAs in terms of resource-error trade-off: not only with 10 error-coefficients and considering 4 MSBs we achieve Average Relative Error (ARE) of 0.6\%, which is better than SIMDive/REALM (0.8\% ARE, considering 3 MSBs), but also the resource footprint of this scheme is 193 LUTs, still lower than SIMDive/REALM counterparts
(see Table \ref{table:results}).

\begin{figure*}[t]
 \centering
  \includegraphics[width=\textwidth]{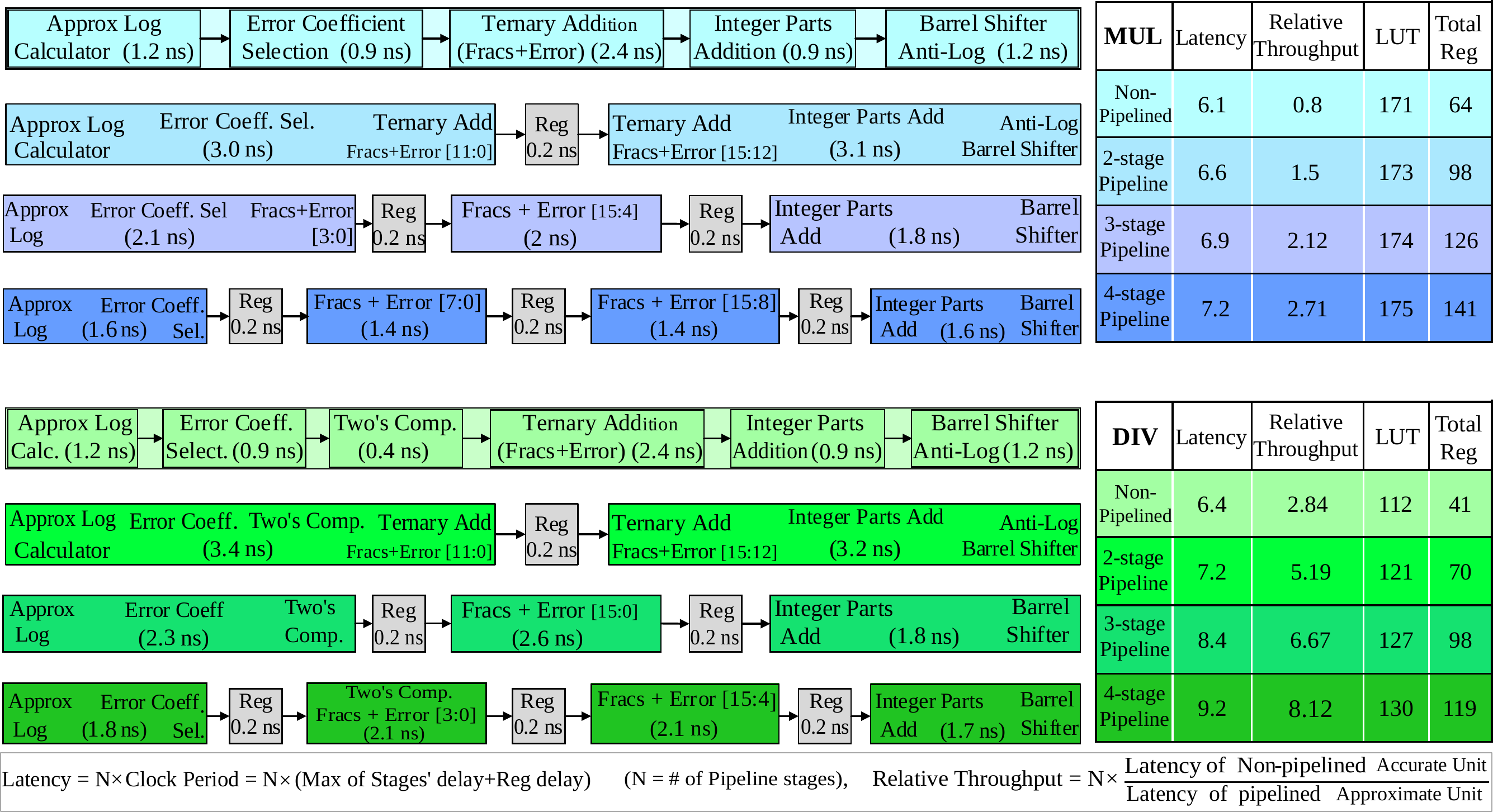}
     \vspace{-0.4cm}
 \caption{Proposed 2-, 3-, and 4-stage pipelining for 16x16 \pname-5 multiplier (top) and \pname-9 16/8 divider (bottom).} \label{fig:Pipelined_structure}
  \vspace{-0.3cm}
\end{figure*} 
\vspace{-0cm}

\subsection{Structure of \pname~multiplier and divider (Non-Pipelined)}
Fig. \ref{fig:proposed_structure} illustrates the structure of the proposed multiplier and divider.
The overall design is designed based on $Res = A \timesdiv B \xrightarrow[\text{Log}]{\text{Approx.}} \widetilde{Log_{Res}} = \widetilde{Log_A} \pm \widetilde{Log_B} \xrightarrow[\text{Anti-Log}]{\text{Approx.}} \widetilde{Res} = 2 ^{\widetilde{Log_{Res}}}$. \normalsize

\ul{Leading-one detection}: To accelerate leading one detection in our FPGA-customized method, this process is calculated
based upon 4-bit LODs (implemented by directly configuring LUTs). In the first step, we prob the presence of bit value `1', simultaneously, in each 4-bit segment of the operands. To this end, one LUT is configured as a logical OR function, applied on 4-bits of each group to reveal whether the segment contains a bit with value ‘1’ (acts as a zero-detection flag). In parallel, another 6-LUT is configured to two 5-LUTs in such a way that it determines the position of leading-one in the 4-bit segment (LOD4-LUT).
Finally, based on the resulting bits from the output of these 6-LUTs, we determine the position of leading one in the most significant group, through the priority logic.
For example, the position of leading one in the 8-bit LOD equals to the concatenation of \{\textit{Location index of most significant segment}, \textit{Leading one position in that segment}\}, e.g.,
the leading one position in "01010101" is \{\{1\},\{10\} =$>$ \{110\}\} in binary.
The first part (\{1\}) is the output of Flag-LUT that has been employed for upper 4-bit segment. The second part (\{10\}) is the result of LOD4-LUTs on the upper segment. A similar method has been also exploited for 16- and 32-bit LODs. For example, in a 16-LOD, if the upper half of the operand is zero, the 16-bit LOD is equal to the lower 8-bit LOD. Else, the position of leading-one is 8+leading-one position in the upper-half 8-LOD. This can be obtained by applying a logical-OR function on the outputs of Flag-LUTs on the 4-bit segments of the upper-half. In LeAp \cite{9045171}, LOD step was orchestrated through an FSM and performed
in, at most, five clock-cycles.
In order to achieve efficient pipelining and minimize the number of registers, the LOD is implemented as combinational logic. Subsequently, the critical path is analyzed in order to achieve balanced partitioning for pipelining.

\ul{Addition of integer parts}: as shown in Fig. \ref{fig:proposed_structure}, each 4-bit addition is fulfilled by one Virtex-7 slice which includes four 6-LUTs and its associated fast carry chains. Together, these implement a Carry Look-Ahead Adder (CLA). Extending the 4-bit addition to 8-bits is easily achievable by connecting the $C_{out}$ from a previous slice to the $C_{in}$ of the next slice.
Recalling Eq. \ref{equation3_1} and Eq. \ref{equation3_2}, division is performed by changing additions to subtractions, through 2's complement modules.

\ul{LUT-optimised ternary addition}: recalling that in LeAp \cite{9045171}, we have proposed error reduction coefficients such that it only depends on the fractional bits (and not the intermediate result of Mitchell Mul/Div, contrary to MBM/INZeD \cite{8493590, 8807077}). On the other hand, LUTs and their associated fast carry chain in Xilinx UNISIM library \cite{Xilinx7SeriesGuide} can be configured to implement a \textit{ternary} adder. To this end, we have manually configured the LUTs and carry chain primitives of the FPGA
in such a way that they implement a ternary adder. This highly suits our error-reduction approach as \emph{enables combining the process of adding error-reduction coefficient with fractional parts with the same resource footprint and in a single step}. In fact, regardless of the ternary adder size and compared to the binary version, only one more bit at MSB position is needed, as frac1$_i$+frac2$_i$+error\_coefficient$_i$+C$_{in}$ (C$_{out}$ from prior bit) may result in 3 bits, requiring another LUT at the end of the chain \cite{9211770}. Moreover, the delay of FPGA primitives is fixed. Therefore, adding error-reduction term at the same time as fractional parts does not impose additional overhead to the design. On the contrary, in REALM \cite{9116315}, MBM \cite{8493590}, and INZeD \cite{8807077} an additional circuit is needed to add the error-reduction term (or half of it) to the original Mitchell's circuits, based on the intermediate summation/subtraction of fractional parts.

It should be noted that to prevent overflow in 2N-by-N bit standard division, the condition for $dividend <2^N \times divisor$ needs to be satisfied \cite{parhami2010computer}. This means that scaling of the divider lies in the range of 2$^0$ to 2$^{N-1}$ (similar to \cite{8807077}). Therefore, only N-1 bits are used from the subtractors for the output and N LSBs from $\widetilde{log}_{dividend}$ is neglected. This also reduces the resource footprint for subtractor and barrel shifter and does not affect the accuracy.
\vspace{-0.2cm}

\subsection{Proposed Pipelined Architectures for Multiplier and Divider}
In order to minimize the latency overhead, the combinational datapath of multiplier and divider should be partitioned
for near uniform latency over the pipelined stages. To achieve this, we have adopted the following steps: first, each stage of the multiplication or division is synthesized in isolation to get an \textit{estimation} of the delay for each stage. Fig. \ref{fig:Pipelined_structure} shows the latency of individual stages in non-pipelined and various pipelined configurations for a 5-coefficient multiplier and divider (the results for all designs are shown in Table \ref{table:results}). Note that after integrating the sub-modules and synthesizing the entire component, the end-to-end latency may change due to the default-optimizations and structural-modification applied by the tool. Based on this analysis, the pipelining registers were inserted in the proper locations of the non-pipelined design. Afterwards, re-synthesis has been performed to assess whether a marginal fine-tuning for the adopted partitioning can result in better end-to-end latency.
It can be inferred from Fig. \ref{fig:Pipelined_structure}, that through the adopted pipelining approach, the latency of stages become similar. Moreover, the delay overhead of a pipeline register is
small,
particularly since the divider is in the critical path in most applications. Moreover, each of the proposed soft-core designs has a different operating frequency and can be utilized w.r.t the different frequency levels offered by a single or various FPGA families. Finally, different pipeline versions enable a spectrum of latency-throughput trade-offs. Therefore, the proper configurations can be loaded w.r.t. the application's requirement.


\section{Results and Discussion
\label{sec:results}}
In this section, we first present the implementation results of the proposed \pname~multipliers and dividers (pipelined and non-pipelined configurations). We have also implemented the compared the proposed \pname~against the following designs: DSP- and pipelined/non-pipelined IP-based accurate counterparts, dynamically truncated DRUM \cite{7372600} and AAXD \cite{8713893}, hierarchical-based AFM multiplier \cite{9045546}, SAADI-EC pipelined divider \cite{8766885}, and Mitchell-based counterparts (SIMDive multiplier and divider \cite{10.1145/3386263.3406907}, MBM multiplier \cite{8493590}, and INZeD divider \cite{8807077}). We compare the designs w.r.t different performance metrics including \textit{Performance (herein throughput) per Watt} which is considered by industry to be the very new Moore’s Law \cite{ArmPeformancePerWatt}. Afterwards, the end-to-end performance gain on three multi-kernel applications
is assessed.

\begin{table*}[!t]
\huge
\centering
\caption{{Accuracy-resource trade-off of accurate/approximate multipliers \& dividers (pipelined/non-pipelined) in FPGA implementation
}}
\vspace{-0.1 cm}
\label{table:results}
\begin{threeparttable}
\def\arraystretch{1.3}
\resizebox{\textwidth}{!}{
\begin{tabular}{|c|c|c|c|c|c|c|c|c|c|c|c?{1mm}c|c|c|c|c|c|c|c|c|c|c|c|} 
\Xhline{5\arrayrulewidth}
\hline
\multicolumn{12}{|c?{1mm}}{\textbf{Multiplier}}   & \multicolumn{12}{c|}{\textbf{Divider}} \\ \Xhline{5\arrayrulewidth}
 \hline
\textbf{8$\times$8 Mul}                      & \textbf{LUT} & \textbf{FF} & \begin{tabular}[c]{@{}c@{}}\textbf{E2E $^\textbf{1}$}\\\textbf{Latency}\\\textbf{(ns)}\end{tabular} & \begin{tabular}[c]{@{}c@{}}\textbf{Rel.$^\textbf{2}$}\\\textbf{Tput}\end{tabular} & \begin{tabular}[c]{@{}c@{}}\textbf{Circuit}\\\textbf{Power}\\\textbf{(mW)}\end{tabular} & \begin{tabular}[c]{@{}c@{}}\textbf{Clk}\\\textbf{Power}\\\textbf{(mW)}\end{tabular} & \begin{tabular}[c]{@{}c@{}}\textbf{Rel. $^\textbf{3}$}\\\textbf{Energy}\\\textbf{per inst}\end{tabular} & \begin{tabular}[c]{@{}c@{}}\textbf{Rel.}\\\textbf{Tput/}\\\textbf{Watt}\end{tabular} & \begin{tabular}[c]{@{}c@{}}\textbf{ARE}\\\textbf{(\%)}\end{tabular} & \begin{tabular}[c]{@{}c@{}}\textbf{PRE}\\\textbf{(\%)}\end{tabular} & \begin{tabular}[c]{@{}c@{}}\textbf{Error}\\\textbf{Bias}\\\textbf{(\%)}\end{tabular} & \textbf{8/4 Div}                     & \textbf{LUT} & \textbf{FF} & \begin{tabular}[c]{@{}c@{}}\textbf{E2E}\\\textbf{Latency}\\\textbf{(ns)}\end{tabular} & \begin{tabular}[c]{@{}c@{}}\textbf{Rel.}\\\textbf{Tput}\end{tabular} & \begin{tabular}[c]{@{}c@{}}\textbf{Circuit}\\\textbf{Power}\\\textbf{(mW)}\end{tabular} & \begin{tabular}[c]{@{}c@{}}\textbf{Clk}\\\textbf{Power}\\\textbf{(mW)}\end{tabular} & \begin{tabular}[c]{@{}c@{}}\textbf{Rel.}\\\textbf{Energy}\\\textbf{per inst}\end{tabular} & \begin{tabular}[c]{@{}c@{}}\textbf{Rel.}\\\textbf{Tput/}\\\textbf{Watt}\end{tabular} & \begin{tabular}[c]{@{}c@{}}\textbf{ARE}\\\textbf{(\%)}\end{tabular} & \begin{tabular}[c]{@{}c@{}}\textbf{PRE}\\\textbf{(\%)}\end{tabular} & \begin{tabular}[c]{@{}c@{}}\textbf{Error}\\\textbf{Bias}\\\textbf{(\%)}\end{tabular}  \\ 
\toprule
\textbf{DSP-based}                    &\begin{tabular}[c]{@{}c@{}} 7 + \\ 1 DSP \end{tabular}   & \cellcolor{lighter_green} 31          & 3.72                                                                                  & 1.97                                                                 & 17.52                                                                                   & \cellcolor{lighter_green} 10.3                                                                                & 1.77                                                                                      & \cellcolor{lighter_green} 1.13                & -            & -            & -                   & \textbf{DSP-based}                   & \begin{tabular}[c]{@{}c@{}}11+ \\2 DSPs \end{tabular}   & 63          & 7.48                                                                                  & 1.44                                                                 & 19.80                                                                                   & \cellcolor{lighter_green} 11.6                                                                                & 2.25                                                                                      & 0.44               & -            & -            & -                    \\ 
\hline
\textbf{Acc IP\_NP $^\textbf{4}$}               & 60           & 48          & \cellcolor{lighter_green} 3.67                                                                                  & 1.0                                                                 & 15.91                                                                                   & -                                                                                   & \cellcolor{lighter_green} 1.0                                                                                      & 1.0                & -            & -            & -                   & \textbf{Acc IP\_NP}              & 51           & 42          & 10.74                                                                                 & 1.0                                                                 & 9.62                                                                                    & -                                                                                   & 1.0                                                                                      & 1.0               & -            & -            & -                    \\ 
\hline
\textbf{Acc IP\_P2}               & 60           & 88          & 4.30                                                                                  & 1.71                                                                 & 23.19                                                                                   & 37.5                                                                                & 2.23                                                                                      & 0.45                & -            & -            & -                   & \textbf{Acc IP\_P2}              & 55           & 59          & 13                                                                                    & 1.67                                                                 & 21.09                                                                                   & 18.4                                                                                & 2.43                                                                                      & 0.41               & -            & -            & -                    \\ 
\hline

\textbf{Acc IP\_P3}               & 60           & 112          & 5.10                                                                                  & \cellcolor{lighter_green} 2.16                                                                 & 47.5                                                                                   & 43                                                                               & 2.64                                                                                      & 0.38                & -            & -            & -                   & \textbf{Acc IP\_P4}              & 69           & 89          & 23.7                                                                                    & 1.81                                                                 & 26.27                                                                                   & 29.1                                                                                & 3.14                                                                                     & 0.32               & -            & -            & -                    \\ 
\hline

\textbf{\pname-3\_NP}   & 57           & 39          & 4.97                                                                                  & 0.74                                                                 & \cellcolor{lighter_green} 12.06                                                                                   & -                                                                                   &  1.03                                                                                      & 0.97                & 1.02         & 6.1         & 0.06                & \textbf{\pname-3\_NP}  & 41           & \cellcolor{lighter_green} 20          & \cellcolor{lighter_green} 5.20                                                                                  & 2.06                                                                 & 9.72                                                                                    & -                                                                                   & 0.48                                                                                      & 2.06               & 0.99         & 5.74         & 0.02                 \\ 
\hline
\textbf{\pname-5\_P2$^\textbf{5}$}  & 62           & 56          & 5.45                                                                                  & 1.35                                                                 & 18.17                                                                                   & 22.9                                                                                & 1.91                                                                                      & 0.52                & 0.91         & \cellcolor{lighter_green} 4.45         & \cellcolor{lighter_green} 0.05                & \textbf{\pname-5\_P2} & 44           & 20          & 5.18                                                                                  & 4.15                                                                 & 14.74                                                                                   & 20.8                                                                                & 0.88                                                                                      & 1.13               & 0.79         & 4.34         & \cellcolor{lighter_green} 0.01                 \\ 
\hline
\textbf{\pname-10\_P3} & 71           & 69          & 6.88                                                                                  & 1.60                                                                 & 24.31                                                                                   & 34.0                                                                                & 2.29                                                                                      & 0.44                &  \cellcolor{lighter_green} 0.64         & 3.69         & 0.05                & \textbf{\pname-9\_P3} & 51           & 20          & 5.34                                                                                  & \cellcolor{lighter_green} 8.05                                                                 & 22.66                                                                                   & 30.6                                                                                & 0.91                                                                                      & 1.47               & \cellcolor{lighter_green} 0.58         & \cellcolor{lighter_green} 3.48         & 0.01                 \\ 
\hline
\textbf{AFM1}                         & 69           & 58          & 5.50                                                                                  & 0.67                                                                 & 18.66                                                                                   & -                                                                                   & 1.76                                                                                      & 0.57                & 0.23         & 16.52        & 0.23                & \textbf{SIMDive-DIV}                 & 44           & 20          & 5.23                                                                                  & 2.09                                                                 & 9.23                                                                                    & -                                                                                   & \cellcolor{lighter_green} 0.45                                                                                      & \cellcolor{lighter_green} 2.20               & 0.77         & 5.20         & 0.01                 \\ 
\hline
\textbf{SIMDive-MUL}                      & 61           & 32          & 5.13                                                                                  & 0.72                                                                 & 16.34                                                                                   & -                                                                                   & 1.44                                                                                      & 0.7                & 0.82         & 4.76         & 0.05                & \textbf{INZeD}                       & 47           & 21          & 6.12                                                                                  & 1.75                                                                 & 11.14                                                                                   & -                                                                                   & 0.65                                                                                      & 1.53               & 2.93         & 9.53         & 0.02                 \\ 
\hline
\textbf{MBM}                          & 64           & 33          & 5.16                                                                                  & 0.71                                                                 & 17.48                                                                                   & -                                                                                   & 1.54                                                                                      & 0.65                & 2.60         & 8.59         & 0.09                & \textbf{Mitchell}                    & \cellcolor{lighter_green} 36           & 20          & \cellcolor{lighter_green} 5.10                                                                                  & 2.11                                                                 & \cellcolor{lighter_green} 8.53                                                                                    & -                                                                                   &  \cellcolor{lighter_green} 0.42                                                                                      & \cellcolor{lighter_green} 2.40               & 3.90         & 13.00        & 3.90                 \\ 
\hline
\textbf{Mitchell}                     & \cellcolor{lighter_green} 51           & 32          & 4.82                                                                                  & 0.76                                                                 & 13.39                                                                                   & -                                                                                   & 1.11                                                                                      & 0.9                & 3.77         & 11.11        & 3.77                & \textbf{SAADI-EC (16)}               & 103          & 42          & 13                                                                                    & 0.84                                                                 & 24.42                                                                                   & -                                                                                   & 2.99                                                                                      & 0.33               & 2.37         & 8.82         & 1.92                 \\ 
\hline
\textbf{DRUM-4}                       & \cellcolor{lighter_green} 53           & 32          & 5.08                                                                                  & 0.72                                                                 & 14.33                                                                                   & -                                                                                   & 1.25                                                                                      & 0.81                & 5.82         & 25.35        & 1.84                & \textbf{AAXD (6/3)}                  & \cellcolor{lighter_green} 38           & 20          & 6.06                                                                                  & 1.77                                                                 & \cellcolor{lighter_green}  9.01                                                                                    & -                                                                                   & 0.52                                                                                      & 1.91               & 2.08         & 100.00       & 1.49                 \\ 
\Xhline{5\arrayrulewidth}
\textbf{16$\times$16 Mul}                    & \textbf{LUT} & \textbf{FF} & \begin{tabular}[c]{@{}c@{}}\textbf{E2E}\\\textbf{Latency}\\\textbf{(ns)}\end{tabular} & \begin{tabular}[c]{@{}c@{}}\textbf{Rel.}\\\textbf{Tput}\end{tabular} & \begin{tabular}[c]{@{}c@{}}\textbf{Circuit}\\\textbf{Power}\\\textbf{(mW)}\end{tabular} & \begin{tabular}[c]{@{}c@{}}\textbf{Clk}\\\textbf{Power}\\\textbf{(mW)}\end{tabular} & \begin{tabular}[c]{@{}c@{}}\textbf{Rel.}\\\textbf{Energy}\\\textbf{per inst}\end{tabular} & \begin{tabular}[c]{@{}c@{}}\textbf{Rel.}\\\textbf{Tput/}\\\textbf{Watt}\end{tabular} & \begin{tabular}[c]{@{}c@{}}\textbf{ARE}\\\textbf{(\%)}\end{tabular} & \begin{tabular}[c]{@{}c@{}}\textbf{PRE}\\\textbf{(\%)}\end{tabular} & \begin{tabular}[c]{@{}c@{}}\textbf{Error}\\\textbf{Bias}\\\textbf{(\%)}\end{tabular} & \textbf{16/8 Div}                    & \textbf{LUT} & \textbf{FF} & \begin{tabular}[c]{@{}c@{}}\textbf{E2E}\\\textbf{Latency}\\\textbf{(ns)}\end{tabular} & \begin{tabular}[c]{@{}c@{}}\textbf{Rel.}\\\textbf{Tput}\end{tabular} & \begin{tabular}[c]{@{}c@{}}\textbf{Circuit}\\\textbf{Power}\\\textbf{(mW)}\end{tabular} & \begin{tabular}[c]{@{}c@{}}\textbf{Clk}\\\textbf{Power}\\\textbf{(mW)}\end{tabular} & \begin{tabular}[c]{@{}c@{}}\textbf{Rel.}\\\textbf{Energy}\\\textbf{per inst}\end{tabular} & \begin{tabular}[c]{@{}c@{}}\textbf{Rel.}\\\textbf{Tput/}\\\textbf{Watt}\end{tabular} & \begin{tabular}[c]{@{}c@{}}\textbf{ARE}\\\textbf{(\%)}\end{tabular} & \begin{tabular}[c]{@{}c@{}}\textbf{PRE}\\\textbf{(\%)}\end{tabular} & \begin{tabular}[c]{@{}c@{}}\textbf{Error}\\\textbf{Bias}\\\textbf{(\%)}\end{tabular}  \\ 
\Xhline{5\arrayrulewidth}
\textbf{DSP-based}                    & \begin{tabular}[c]{@{}c@{}}8 +\\ 1 DSP  \end{tabular} & \cellcolor{lighter_green} 32          & \cellcolor{lighter_green} 4.11                                                                                  & 1.19                                                                 & \cellcolor{lighter_green} 17.48                                                                                   & \cellcolor{lighter_green} 9.8                                                                                 & \cellcolor{lighter_green} 0.48                                                                                      &  \cellcolor{lighter_green} 2.08                & -            & -            & -                   & \textbf{DSP-based}                   & \begin{tabular}[c]{@{}c@{}}197+\\7DSP  \end{tabular}   & 131         & 10.36                                                                                 & 1.76                                                                 & 51.62                                                                                 & 25.1                                                                                      & 2.34   & 0.43            & -            & -            & -                    \\ 
\hline
\textbf{Acc IP\_NP}               & 287          & 64          & 4.88                                                                                  & 1.0                                                                 & 47.81                                                                                   & -                                                                                   & 1.0                                                                                      & 1.0                & -            & -            & -                   & \textbf{Acc IP\_NP}              & 169          & 76          & 18.23                                                                                 & 1.0                                                                 & \cellcolor{lighter_green}  17.97                                                                                   & -                                                                                   & 1.0                                                                                      & 1.0               & -            & -            & -                    \\ 
\hline
\textbf{Acc IP\_P2}               & 249          & 176         & 6.14                                                                                  & 1.60                                                                 & 64.86                                                                                   & 65.3                                                                                & 1.71                                                                                      & 0.59                & -            & -            & -                   & \textbf{Acc IP\_P2}              & 175          & 104         & 19.59                                                                                 & 1.86                                                                 & 38.82                                                                                   & \cellcolor{lighter_green} 12.3                                                                                & 1.47                                                                                      & 0.68               & -            & -            & -                    \\ 
\hline
\textbf{Acc IP\_P3}               & 249          & 245         & 8.88                                                                                  & 1.65                                                                 & 94.49                                                                                   & 75.0                                                                                & 2.15                                                                                      & 0.46                & -            & -            & -                   & \textbf{Acc IP\_P4}              & 181          & 168         & 20.09                                                                                 & 3.63                                                                 & 56.21                                                                                   & 24.5                                                                                & 1.68                                                                                      & 0.6               & -            & -            & -                    \\ 
\hline
\textbf{Acc IP\_P4}               & 249          & 343         & 9.60                                                                                  & 2.03                                                                 & 150.73                                                                                  & 130.0                                                                               & 2.89                                                                                     & 0.35                & -            & -            & -                   & \textbf{\pname-3\_NP}  & \cellcolor{lighter_green} 112          & 41          & 6.38                                                                                  & 2.98                                                                 &   18.67                                                                                   & -                                                                                   & \cellcolor{lighter_green} 0.34                                                                                      & \cellcolor{lighter_green} 2.98               & 1.02         & 5.74         & 0.02                 \\ 
\hline
\textbf{\pname-3\_NP} & \cellcolor{lighter_green} 168          & 64          & 5.90                                                                                  & 0.83                                                                 & 31.43                                                                                   & -                                                                                   & 0.86                                                                                      & 1.17                & 1.03         & 6.1         & 0.06                & \textbf{\pname-5\_P2} & 121          & 70          & 7.07                                                                                  & 5.16                                                                 & 27.77                                                                                   & 19.5                                                                                & 0.49                                                                                      & 2.04               & 0.79         & 4.34         & 0.01                 \\ 
\hline
\textbf{\pname-3\_P2}  & 169          & 98          & 6.11                                                                                  & 1.52                                                                 & 47.35                                                                                   & 40.7                                                                                & 1.21                                                                                      & 0.85                & 1.03         & 6.1         & 0.06                & \textbf{\pname-9\_P3} & 127          & 98          & 8.35                                                                                  & 6.62                                                                 & 30.29                                                                                   & 24.3                                                                                 & 0.44                                                                                      & 2.28               & \cellcolor{lighter_green} 0.58         & \cellcolor{lighter_green} 3.48         & \cellcolor{lighter_green} 0.01                 \\ 
\hline
\textbf{\pname-5\_P3}  & 177          & 126         & 6.87                                                                                  & 2.25                                                                 & 75.35                                                                                   & 58.2                                                                                & 1.27                                                                                      & 0.81                & 0.93         & \cellcolor{lighter_green} 4.45         & 0.05                & \textbf{\pname-9\_P4} & 130          & 119         & 9.2                                                                                  & \cellcolor{lighter_green} 8.01                                                                 & 34.68                                                                                   & 27.7                                                                                & 0.42                                                                                      & 2.40               & 0.58         & 3.48         & 0.01                 \\ 
\hline
\textbf{\pname-10\_P4} & 193          & 141         & 7.25                                                                                  & \cellcolor{lighter_green} 2.52                                                                 & 84.75                                                                                   & 87.4                                                                                & 1.46                                                                                      & 0.7                & \cellcolor{lighter_green} 0.56         & 3.69         & 0.23                & \textbf{SIMDive-DIV}                 & 143          & 64          & \cellcolor{lighter_green} 5.68                                                                                  & 3.28                                                                 & 23.84                                                                                   & -                                                                                   & 0.39                                                                                      &  2.57               & 0.78         & 5.20         & 0.01                 \\ 
\hline
\textbf{AFM1}                         & 261          & 66          & 7.32                                                                                  & 0.67                                                                 & 44.78                                                                                   & -                                                                                   & 1.41                                                                                      & 0.71               & 1.34         & 17.80        & 1.34                & \textbf{INZeD}                       & 165          & \cellcolor{lighter_green}41          & 6.28                                                                                  & 2.90                                                                 & 27.50                                                                                   & -                                                                                   & 0.51                                                                                      & 1.97               & 2.93         & 9.54         & 0.02                 \\ 
\hline
\textbf{SIMDive-MUL}                      & 216          & 64          & 5.95                                                                                  & 0.82                                                                 & 37.06                                                                                   & -                                                                                   & 0.95                                                                                      & 1.06                & 0.82         & 4.90         & 0.05                & \textbf{Mitchell}                    & \cellcolor{lighter_green} 106          & 64          & \cellcolor{lighter_green} 5.56                                                                                  & 3.39                                                                 & \cellcolor{lighter_green}  17.34                                                                                   & -                                                                                   &  \cellcolor{lighter_green} 0.32                                                                                      & \cellcolor{lighter_green} 3.11               & 4.11         & 13.00        & 4.11                 \\ 
\hline
\textbf{MBM}                          & 204          & 65          & 6.59                                                                                  & 0.74                                                                 & 35.34                                                                                   & -                                                                                   & 1.0                                                                                      & 1.0                & 2.63         & 8.83         & 0.09                & \textbf{SAADI-EC (16)}               & 342          & 126         & 25.70                                                                                 & 0.71                                                                 & 57.01                                                                                   & -                                                                                   & 4.31                                                                                      & 0.23               & 2.14         & 8.82         & 1.76                 \\ 
\hline
\textbf{Mitchell}                     & \cellcolor{lighter_green} 167          & 64          & 5.51                                                                                  & 0.99                                                                 & 31.46                                                                                   & -                                                                                   & 0.64                                                                                      & 1.56                & 3.85         & 11.11        & 3.85                & \textbf{AAXD (8/4)}                  & 151          & 155         & 12.51                                                                                 & 1.46                                                                 & 25.17                                                                                   & -                                                                                   & 0.93                                                                                      &   1.08               & 2.99         & 100          & 0.90                 \\ 
\hline
\textbf{DRUM-6}                       & 233          & 64          & 5.34                                                                                  & 0.91                                                                 & 38.43                                                                                   & -                                                                                   & 0.88                                                                                      & 1.14                & 1.47         & 6.31         & \cellcolor{lighter_green} 0.04                & \textbf{AAXD (12/6)}                 & 207          & 233         & 21.26                                                                                 & 0.86                                                                 & 34.51                                                                                   & -                                                                                   & 2.16                                                                                      & 0.46               & 0.74         & 100          & 0.30                 \\ 
\Xhline{5\arrayrulewidth}
\textbf{32$\times$32 Mul}                    & \textbf{LUT} & \textbf{FF} & \begin{tabular}[c]{@{}c@{}}\textbf{E2E}\\\textbf{Latency}\\\textbf{(ns)}\end{tabular} & \begin{tabular}[c]{@{}c@{}}\textbf{Rel.}\\\textbf{Tput}\end{tabular} & \begin{tabular}[c]{@{}c@{}}\textbf{Circuit}\\\textbf{Power}\\\textbf{(mW)}\end{tabular} & \begin{tabular}[c]{@{}c@{}}\textbf{Clk}\\\textbf{Power}\\\textbf{(mW)}\end{tabular} & \begin{tabular}[c]{@{}c@{}}\textbf{Rel.}\\\textbf{Energy}\\\textbf{per inst}\end{tabular} & \begin{tabular}[c]{@{}c@{}}\textbf{Rel.}\\\textbf{Tput/}\\\textbf{Watt}\end{tabular} & \begin{tabular}[c]{@{}c@{}}\textbf{ARE}\\\textbf{(\%)}\end{tabular} & \begin{tabular}[c]{@{}c@{}}\textbf{PRE}\\\textbf{(\%)}\end{tabular} & \begin{tabular}[c]{@{}c@{}}\textbf{Error}\\\textbf{Bias}\\\textbf{(\%)}\end{tabular} & \textbf{32/16 Div}                   & \textbf{LUT} & \textbf{FF} & \begin{tabular}[c]{@{}c@{}}\textbf{E2E}\\\textbf{Latency}\\\textbf{(ns)}\end{tabular} & \begin{tabular}[c]{@{}c@{}}\textbf{Rel.}\\\textbf{Tput}\end{tabular} & \begin{tabular}[c]{@{}c@{}}\textbf{Circuit}\\\textbf{Power}\\\textbf{(mW)}\end{tabular} & \begin{tabular}[c]{@{}c@{}}\textbf{Clk}\\\textbf{Power}\\\textbf{(mW)}\end{tabular} & \begin{tabular}[c]{@{}c@{}}\textbf{Rel.}\\\textbf{Energy}\\\textbf{per inst}\end{tabular} & \begin{tabular}[c]{@{}c@{}}\textbf{Rel.}\\\textbf{Tput/}\\\textbf{Watt}\end{tabular} & \begin{tabular}[c]{@{}c@{}}\textbf{ARE}\\\textbf{(\%)}\end{tabular} & \begin{tabular}[c]{@{}c@{}}\textbf{PRE}\\\textbf{(\%)}\end{tabular} & \begin{tabular}[c]{@{}c@{}}\textbf{Error}\\\textbf{Bias}\\\textbf{(\%)}\end{tabular}  \\ 
\Xhline{5\arrayrulewidth}
\textbf{DSP-based}                    & \begin{tabular}[c]{@{}c@{}} 18 + \\ 4 DSPs  \end{tabular}& \cellcolor{lighter_green} 75          & 6.93                                                                                  & 1.04                                                                 & \cellcolor{lighter_green} 35.48                                                                                   & \cellcolor{lighter_green} 14.8                                                                                & \cellcolor{lighter_green} 0.47                                                                                      & \cellcolor{lighter_green} 2.28               & -            & -            & -                   & \textbf{DSP-based}                   & \begin{tabular}[c]{@{}c@{}}359+\\9DSP  \end{tabular}   & 228         & 11.65                                                                                 & 3.62                                                                 & 92.91                                                                                   & 28.7                                                                                & 0.98                                                                                      & 1.02               & -            & -            & -                    \\ 
\hline
\textbf{Acc IP\_NP}               & 1012         & 128         & 6.69                                                                                  & 1.0                                                                 & 110.56                                                                                   & -                                                                                   & 1.0                                                                                      & 1.0                & -            & -            & -                   & \textbf{Acc IP\_NP}              & 597          & 139         & 42.24                                                                                 & 1.0                                                                 & \cellcolor{lighter_green}  34.16                                                                                   & -                                                                                   & 1.0                                                                                      & 1.0               & -            & -            & -                    \\ 
\hline
\textbf{Acc IP\_P2}               & 1012         & 291         & 7.82                                                                                  & 1.71                                                                 & 154.26                                                                                   & 92.8                                                                                & 1.31                                                                                      & 0.76                & -            & -            & -                   & \textbf{Acc IP\_P2}              & 607          & 208         & 47.48                                                                                 & 1.78                                                                 & 74.55                                                                                   & \cellcolor{lighter_green} 10.4                                                                                & 1.4                                                                                      & 0.74               & -            & -            & -                    \\ 
\hline
\textbf{Acc IP\_P3}               & 1012         & 835         & 9.74                                                                                  & 2.06                                                                 & 218.32                                                                                  & 146.6                                                                               & 1.6                                                                                      & 0.63                & -            & -            & -                   & \textbf{Acc IP\_P4}              & 607          & 339         & 56.95                                                                                 & 2.97                                                                 & 111.04                                                                                  & 21.9                                                                                & 1.75                                                                                      & 0.76               & -            & -            & -                    \\ 
\hline
\textbf{Acc IP\_P4}               & 1014         & 1119        & 11.73                                                                                 & 2.28                                                                 & 330.64                                                                                  & 250.8                                                                               & 2.31                                                                                     & 0.43                & -            & -            & -                   & \textbf{\pname-3\_NP}  & 378          & 85          & \cellcolor{lighter_green} 6.70                                                                                  &  6.30                                                                 & 42.70                                                                                   & -                                                                                   & \cellcolor{lighter_green} 0.19                                                                                      & 5.04               & 1.04         & 5.74         & 0.02                 \\ 
\hline
\textbf{\pname-3\_NP} & \cellcolor{lighter_green} 434          & 128         & \cellcolor{lighter_green} 6.30                                                                                  & 1.06                                                                 & 90.08                                                                                   & -                                                                                   & 0.77                                                                                      & 1.3                & 1.05         & 6.1         & 0.07                & \textbf{\pname-5\_P2} & 389          & 128         & 8.43                                                                                  & 10.02                                                                & 49.07                                                                                   & 16.5                                                                                & 0.2                                                                                      & 5.22               & 0.79         & 4.34         & \cellcolor{lighter_green} 0.01                 \\ 
\hline
\textbf{\pname-3\_P2}  & 450          & 184         & 6.74                                                                                  & 1.98                                                                 & 131.22                                                                                   & 62.7                                                                                & 0.88                                                                                      & 1.13              & 1.05         & 6.1         & 0.07                & \textbf{\pname-9\_P3} & 399          & 173         & 9.86                                                                                  & 12                                                                   & 59.20                                                                                   & 20.2                                                                                & 0.2                                                                                      & 5.24               & \cellcolor{lighter_green} 0.61         & \cellcolor{lighter_green} 3.48         & 0.01                 \\ 
\hline
\textbf{\pname-5\_P3}  & 462          & 242         & 7.38                                                                                  & 2.72                                                                 & 181.53                                                                                  & 95.1                                                                                & 0.92                                                                                      & 1.09               & 0.95         & \cellcolor{lighter_green} 4.45         & \cellcolor{lighter_green} 0.05                & \textbf{\pname-9\_P4} & 417          & 213         & 11.10                                                                                 & \cellcolor{lighter_green} 15.22                                                                & 67.59                                                                                   & 26.7                                                                                & 0.2                                                                                     & \cellcolor{lighter_green}  5.53               & 0.61         & 3.48         & 0.01                 \\ 
\hline
\textbf{\pname-10\_P4} & 490          & 276         & 8.04                                                                                 & \cellcolor{lighter_green} 3.33                                                                & 253.12                                                                                  & 124.2                                                                               & 1.03                                                                                      & 0.97                & \cellcolor{lighter_green} 0.58         & 3.64         & 0.06                & \textbf{SIMDive-DIV}                 & 381          & \cellcolor{lighter_green} 80          & 6.84                                                                                  & 6.18                                                                 & 43.04                                                                                   & -                                                                                   & 0.2                                                                                   & 4.90               & 0.81         & 5.16         & 0.02                 \\ 
\hline
\textbf{AFM1}                         & 995          & 317         & 10.76                                                                                 & 0.62                                                                 & 119.63                                                                                  & -                                                                                   & 1.74                                                                                     & 0.57                  & 2.88         & 22.40        & 2.88                & \textbf{INZeD}                       & 422          & 81          & 8.15                                                                                  & 5.18                                                                 & 47.67                                                                                   & -                                                                                   & 0.27                                                                                      & 3.71               & 2.96         & 9.47         & 0.03                 \\ 
\hline
\textbf{SIMDive-MUL}                      & 521          & 128         & 6.88                                                                                  & 0.97                                                                 & 83.50                                                                                   & -                                                                                   & 0.78                                                                                     & 1.29                  & 0.91         & 4.72         & 0.05                & \textbf{Mitchell}                    & \cellcolor{lighter_green} 349          & 80          & \cellcolor{lighter_green} 6.21                                                                                  & 6.91                                                                 & 39.42                                                                                   & -                                                                                   & \cellcolor{lighter_green} 0.17                                                                                      & \cellcolor{lighter_green} 5.99               & 4.19         & 13.00        & 4.19                 \\ 
\hline
\textbf{MBM}                          & 533          & 129         & 7.51                                                                                  & 0.89                                                                 & 89.94                                                                                   & -                                                                                   & 0.91                                                                                      & 1.09                  & 2.69         & 8.74         & 0.10                & \textbf{SAADI-EC (16)}               & 822          & 228         & 51.60                                                                                 & 0.82                                                                 & 92.86                                                                                   & -                                                                                   & 3.32                                                                                      & 0.30               & 2.33         & 9.04         & 1.85                 \\ 
\hline
\textbf{Mitchell}                     & \cellcolor{lighter_green} 428         & 128         & \cellcolor{lighter_green} 6.23                                                                                  & 1.12                                                                 & 60.25                                                                                   & -                                                                                   & 0.55                                                                                     & 1.83                  & 3.91         & 11.11        & 3.91                & \textbf{AAXD (8/4)}                  & \cellcolor{lighter_green} 361          & 278         & 24.66                                                                                 & 1.71                                                                 & 40.78                                                                                   & -                                                                                   & 0.70                                                                                      & 1.43               & 3.04         & 100          & 1.10                 \\ 
\hline
\textbf{DRUM-6}                       & 616          & 128         & 6.35                                                                                  & 1.05                                                                 & 88.85                                                                                   & -                                                                                   & 0.76                                                                                      & 1.31                  & 1.53         & 5.88         & 0.05                & \textbf{AAXD (12/6)}                 & 513          & 505         & 37.20                                                                                 & 1.14                                                                 & 57.95                                                                                   & -                                                                                   & 1.49                                                                                      & 0.67               & 0.79         & 100          & 0.35                 \\
\hline
\end{tabular}
}
\begin{tablenotes}\footnotesize
\item $^1$End-to-end\;\;\;  $^2$Throughput\;\;\;  $^3$Energy per instruction = total dynamic power\,$\times$\,clock period\;\;\;
$^4$Non-Pipelined\;\;\;
$^5$\,5-Coeff. 2-stage pipelined 
\end{tablenotes}
\end{threeparttable}
\vspace{-0.4 cm}
\end{table*}


\subsection{\pname~versus SoA multipliers \& dividers (circuit-level evaluation)}
\ul{Experimental Setup}: to evaluate performance metrics, all the designs are developed in Verilog HDL, synthesized, and implemented in Xilinx Vivado 2019.2 for the Virtex-7 FPGA. To ensure scalability of multipliers and dividers, they are compared for precisions of 8-, 16-, and 32-bit. Area and latency are collected from Vivado reports. Power and energy dissipations\footnote{Similar to prior works, only dynamic power is reported from Xilinx Vivado,\;as\;the\;static\;\,power\;analysis\;of\;Vivado\;is\;for\;the\;\,entire\;FPGA\;\cite{9344673, 9072581}.} are obtained through simulations in Xilinx Power Estimator (XPE) over 100 million inputs, uniformly distributed in a random order over the whole input interval. For a precise measurement, we used the clock gating command to prevent superfluous switching activity in unused resources.
To assess accuracy metrics for different partitioning, the behavioral structure of multipliers and dividers are developed in C++. To calculate the average of absolute relative error (ARE), peak absolute relative error (PRE), and error bias in 8- and 16-bit designs, exhaustive testing is performed. For 32-bit error characterization, $2^{32},\sim4.3$ billion input pairs, uniformly distributed over the whole 32-bit interval have been evaluated in Monte Carlo simulations\footnote{Simulated\;on\;Rack\;Server:\,Intel\;Xeon\;E5$\mbox{-}$2667\;@3.2\;GHz,\,512\,GB\;RAM
}.
Post-implementation results are summarized in Table \ref{table:results}.
The following conclusions can be made based on the results:

\begin{itemize}[leftmargin=*]

\item \textbf{\pname~multipliers and dividers versus DSP-based and accurate Vivado IPs}:
the results corroborate prior studies \cite{8532582, 8332524} stating that DSPs are able to be efficiently-utilized, only for large bit-width precision. The reason behind this degraded performance is that DSP48E1 hosts a 25$\times$18 hard-wired multiplier and is not optimized for smaller multipliers. This has been the main motive that FPGA designers have recommended utilizing of soft multipliers should be used for implementing lower bit-widths \cite{8532582, 8332524, XilinxDSPCore, XilinxDivIpCore}. In particular, compared to LUT-based implementation, DSP-based dividers are less energy-efficient, for both 8- and 16-bit precision.
Targeting higher-order precision of 32-bit, the proposed \pname~multiplier and divider have lower latency than DSP-based counterparts. In terms of energy-efficiency (relative energy/instruction), DSPs are better than \pname, only for multiplication. In contrast, \pname~dividers are significantly better than DSP-based implementation (especially in 32-bit precision).
Compared to LUT-based IPs, the results exhibit the efficiency of \pname~divider in all performance metrics, when targeting division of any precision. In case of multiplication, although accurate IP renders better performance in 8-bit over \pname, the area- and/or energy-saving of \pname~over accurate Vivado IP becomes substantial in 16-bit. Moreover, the latency of \pname~also becomes smaller when targeting 32-bit.
Overall, the improvements become more pronounced, when architectures are implemented for higher bit-width. This is due to: first, transforming the 2D array-based structure of Mul/Div to 1D Add/Sub through Mitchell's algorithms, as discussed earlier. Second, the cost of error coefficients gets amortized in higher-order designs.


\item \textbf{\pname~multipliers and dividers render better performance over hierarchical structures}:
comparing Mitchell-based \pname~with modular counterpart AFM (structured by incorporating smaller inexact instances) demonstrates three points: first, thanks to transforming the 2D array structure of multiplication to 1D addition in the logarithmic representation, the area of Mitchell-based unit grows by the factor of $\sim$2.6, less aggressive compared to $\sim$3.7 for the array-based counterparts. This further accentuates the efficiency of \pname, particularly in larger bit-width. Second, comparing the results of 8- versus 16-bit asserts that approximation applied on hierarchical approaches is beneficial in accuracy-resource trade-off, only when it is done from scratch for each multiplier size, otherwise accumulated error in larger designs significantly sacrifices output accuracy to achieve resource efficiency. On the other hand, accuracy metrics in Mitchell-based designs do not undergo notable changes. Finally, as already discussed in earlier work \cite{9045171}, some modular architectures sacrifice delay for LUT saving \cite{10.1145/3195970.3195996}.

\item \textbf{\pname~designs establish better resource-accuracy trade-off than leading-one based truncated counterparts}:
the higher accuracy levels of DRUM and AAXD come at the cost of higher resource consumption than logarithmic counterparts (due to using an \textit{accurate} core unit). Furthermore, employing an accurate instance of a divider still results in a long latency, multiple times that of a same-size multiplier. \pname, on the other hand, has reduced the high latency of accurate divider, nearly to latency of its same-size multiplier. Overall, our proposed architectures yield better resource-accuracy trade-off, especially in higher bit-width. 
Finally, there are many cases with an error near or equal to 100\% in the truncation-based AAXD divider. Such high error cases can result in false positive peaks in heartbeat and corner detection, as will be discussed later.

\item \textbf{The proposed error-reduction outperforms other Mitchell-based architectures}:
the error-reduction strategy of \pname~surpasses SIMDive/REALM in three aspects. First, the proposed error-reduction scheme achieves a higher accuracy level, even with fewer coefficients/LUTs. This is due to the better partitioning scheme of \pname, while it still considers 4 MSBs of fractional parts.
Second, the exponential increase in the number of coefficients (256 for 4 MSBs) of REALM/SIMDive not only poses a noticeable resource penalty, but also would nullify the gain, when realized in a LUT-based implementation. 
Overall, \pname~enables a more cost-effective strategy for partitioning the squarish zone.
Third, in addition to error-reduction strategy, \pname~can achieve a higher throughput, energy, and throughput per Watt, compared to SIMDive, in non-pipeline configuration (except for 8- and 16-bit division). Albeit, as noted previously, the higher throughput enabled by pipelining \pname~comes at the cost of increased throughput per Watt, when compared to SIMDive.

Please note although comparing the SIMD and pipelined architectures is not in the scope of this paper, herein some of the key differences between \pname~and SIMDive are discussed. Supporting higher throughput through SIMDive has added to the complexity of the sub-modules. For example, 4$\times$3=12 bit is used for LOD in 32-bit SIMDive or 2- and 4-MUX units are used to select the functionality and sub-word length in the intermediate adders. Moreover, integration of a multiplier and divider into a hybrid design marginally affects the complexity of the circuit through including 2-MUX units to select the mode in sub-modules. Finally, in order to support simultaneous scalings for sub-word length in SIMD mode, the LOD, error-reduction, and final shifter become more complex. For example, in a 32-bit LOD, instead of 5 bit for the SISD mode, 4$\times$3=12 bit is used to also support simultaneous 8-bit leading one detection for the SIMD mode (the complexity overhead is also posed to integer/fractional part adder).

\item \textbf{Comparing latency, throughput and throughput per Watt in pipelined and non-pipelined designs}: overall, operating at a higher frequency and producing one-operation-per-cycle through pipelining has resulted in additional flip-flops, increased end-to-end latency, and higher dynamic power dissipation.
However, pipelining enables significant improvement in throughput. In fact, higher throughput enabled by increasing the number of pipeline stages comes at the cost of lower \textit{Throughput per Watt} as well. 
This descending trend is mainly due to the increase in the number of FFs and end-to-end latency in such designs.
In particular, the analysis of latency and throughput per Watt highlights the efficacy of pipelining for \pname~circuits from three perspectives.
First, comparing accurate- and \pname-pipelined designs demonstrates that, even with increasing the number of pipelining stages, the end-to-end latency for each x-stage based \pname~remains smaller than its x-stage based accurate counterpart (except for 8-bit multiplication). 
Second, the x-stage based pipelined \pname~enables higher throughput per Watt versus its x-stage based \textit{accurate} counterpart.
Third, while increasing the number of pipeline stages results in a descending trend in throughput per Watt for \pname~multipliers and accurate IPs (both multiplier and divider), this trend is ascending for \pname~dividers. This means that increasing the pipeline depth is highly beneficial for dividers. Moreover, the \textit{relative throughput per Watt} ratio for \pname~dividers are higher than 1, meaning that pipelined \pname~dividers achieve a better throughput per Watt than their accurate counterparts (even when compared to the non-pipelined mode).
On the other hand, as discussed in Section \ref{sec:related} the reciprocal-based dividers are not suited for LUT-based platforms. In fact, the poor performance of pipelined SAADI-EC is due to two reasons: first, its datapath is divided into three non-uniform stages (normalizing block, multiplier, and error-correction accumulator based on Taylor iterations). Second, generating the reciprocal of the divisor (even through utilizing reciprocal IP of Vivado, as adopted herein) is a costly operation for LUT-based designs.\\
It is worth noting that Xilinx Vivado offers different implementation strategies to reduce the dynamic power, including Power\_DefaultOpt, Power\_ExploreArea (in which sequential resources are combined) and system-level power reduction techniques such as voltage scaling \cite{XilinxPowerOpt}. Overall, it is reported that such optimizations can enable up to 30\% improvement in dynamic power \cite{XilinxPowerOpt}. However, considering the trade-off between area/power and performance, exploiting such techniques will increase the critical path delay. As pipelining has increased the critical path delay by itself, we have refrained from applying further directives, but users can utilize such options w.r.t. their constraints.

\begin{figure}[t]
 \centering
  \includegraphics[width=0.51\textwidth]{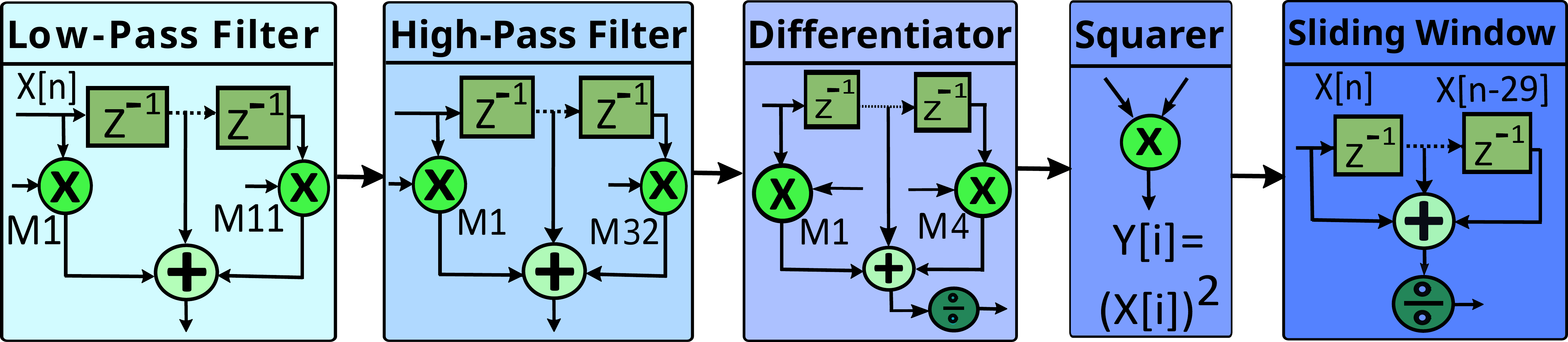}
     \vspace{-0.5cm}
 \caption{The structure of Pan Tompkins QRS detection application \cite{10.1145/3486616} }\label{fig:pan_tompkins}
  \vspace{-0.1cm}
\end{figure} 

\begin{figure}[t]
 \centering
  \includegraphics[width=0.5\textwidth]{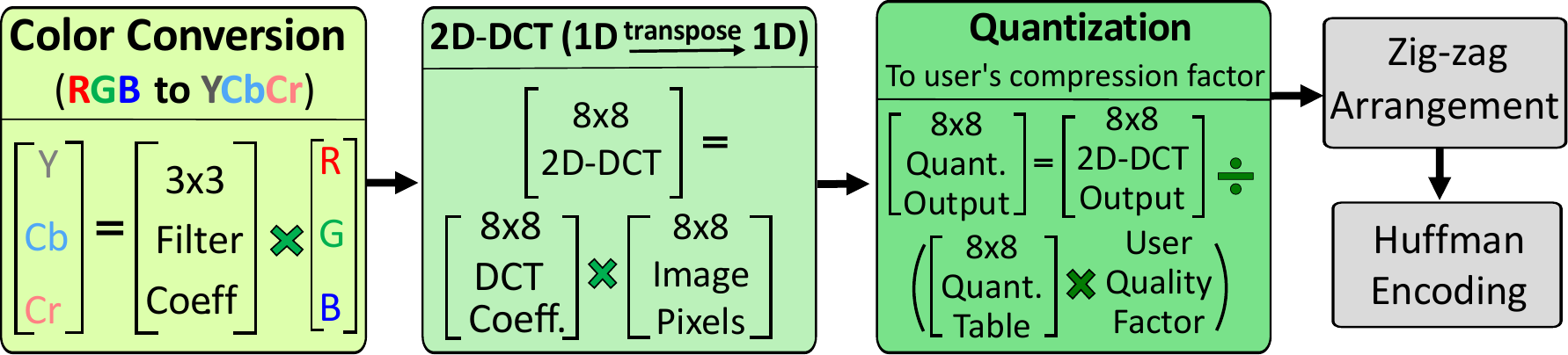}
     \vspace{-.5cm}
 \caption{The structure of JPEG Compression application \cite{10.1145/3486616}}\label{fig:jpeg}
  \vspace{-0.1cm}
\end{figure} 

\begin{figure}[t]
 \centering
  \includegraphics[width=0.46\textwidth]{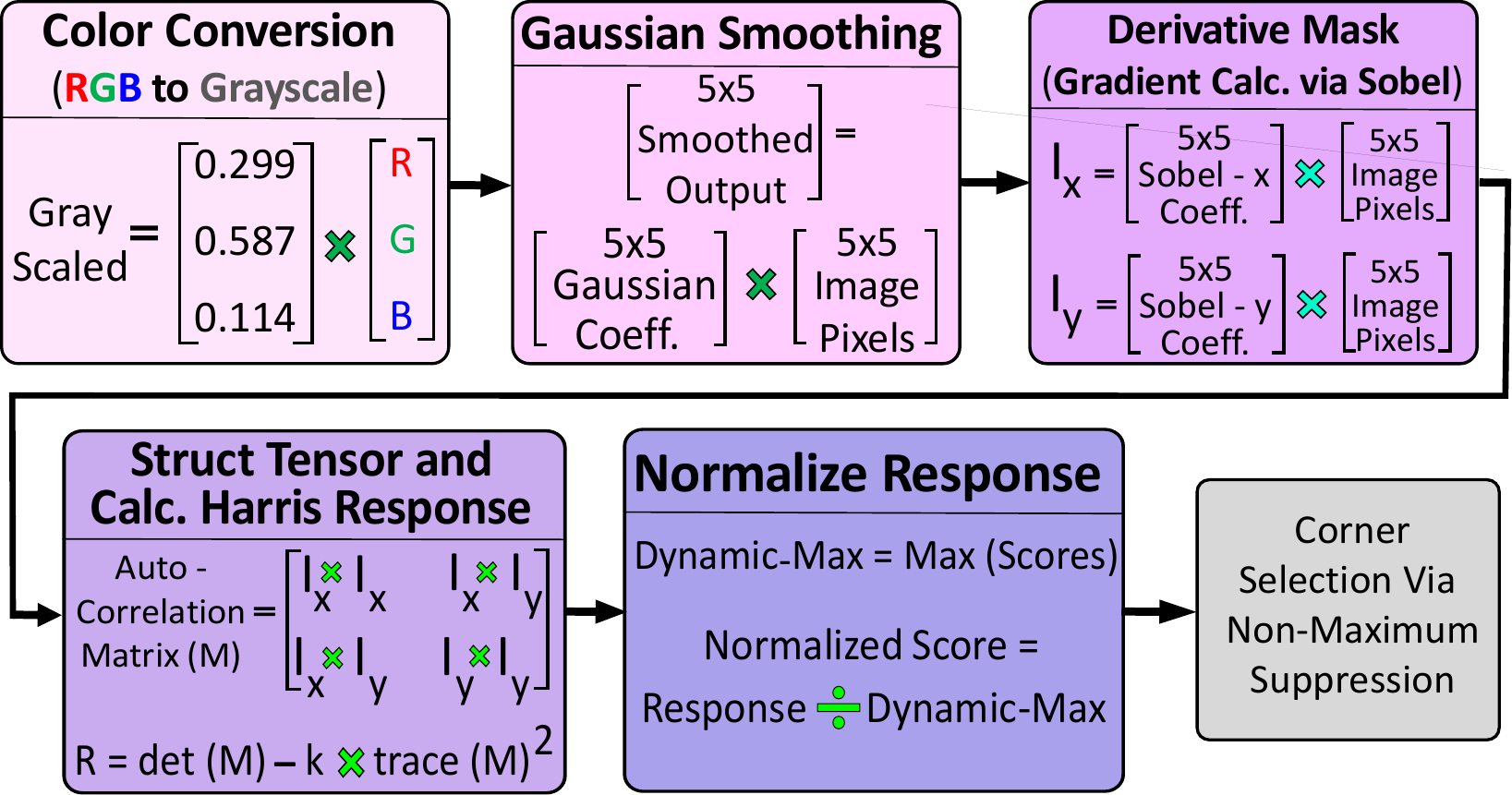}
     \vspace{-.1cm}
 \caption{The structure of Harris Corner Detection application \cite{10.1145/3486616}}\label{fig:harris}
  \vspace{-0.4cm}
\end{figure} 

\begin{figure*}[!h]
    \centering

            \vspace{-0.3cm}
 \subfloat[Accurate multiplication and division (PSNR = 30.9)\label{sub_fig5}]{%
      \includegraphics[width=0.47\textwidth]{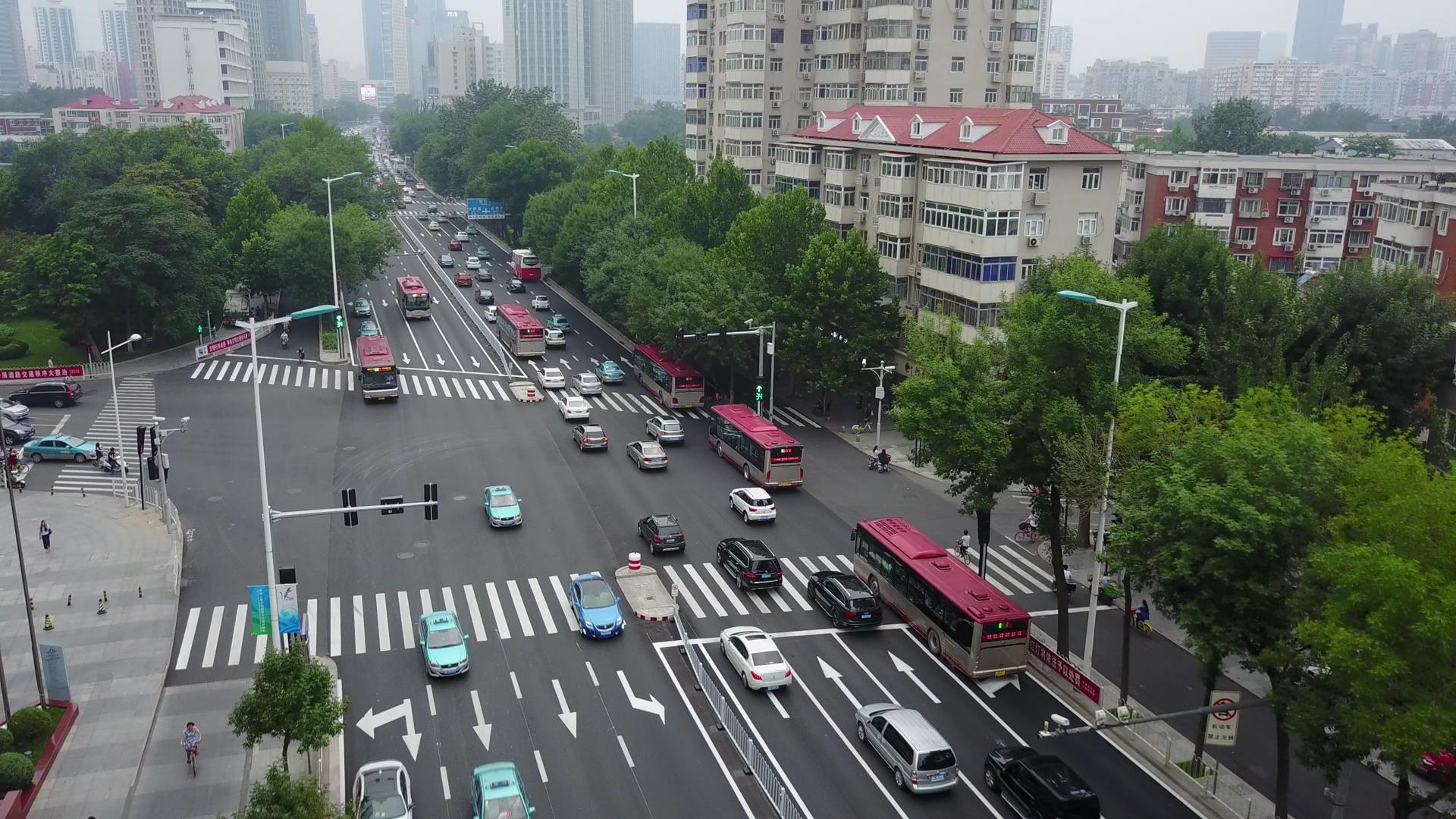}}
    \hfill
  \subfloat[\pname~multiplier-10 and divider-9 (PSNR = 28.7)\label{sub_fig6}]{%
        \includegraphics[width=0.47\textwidth]{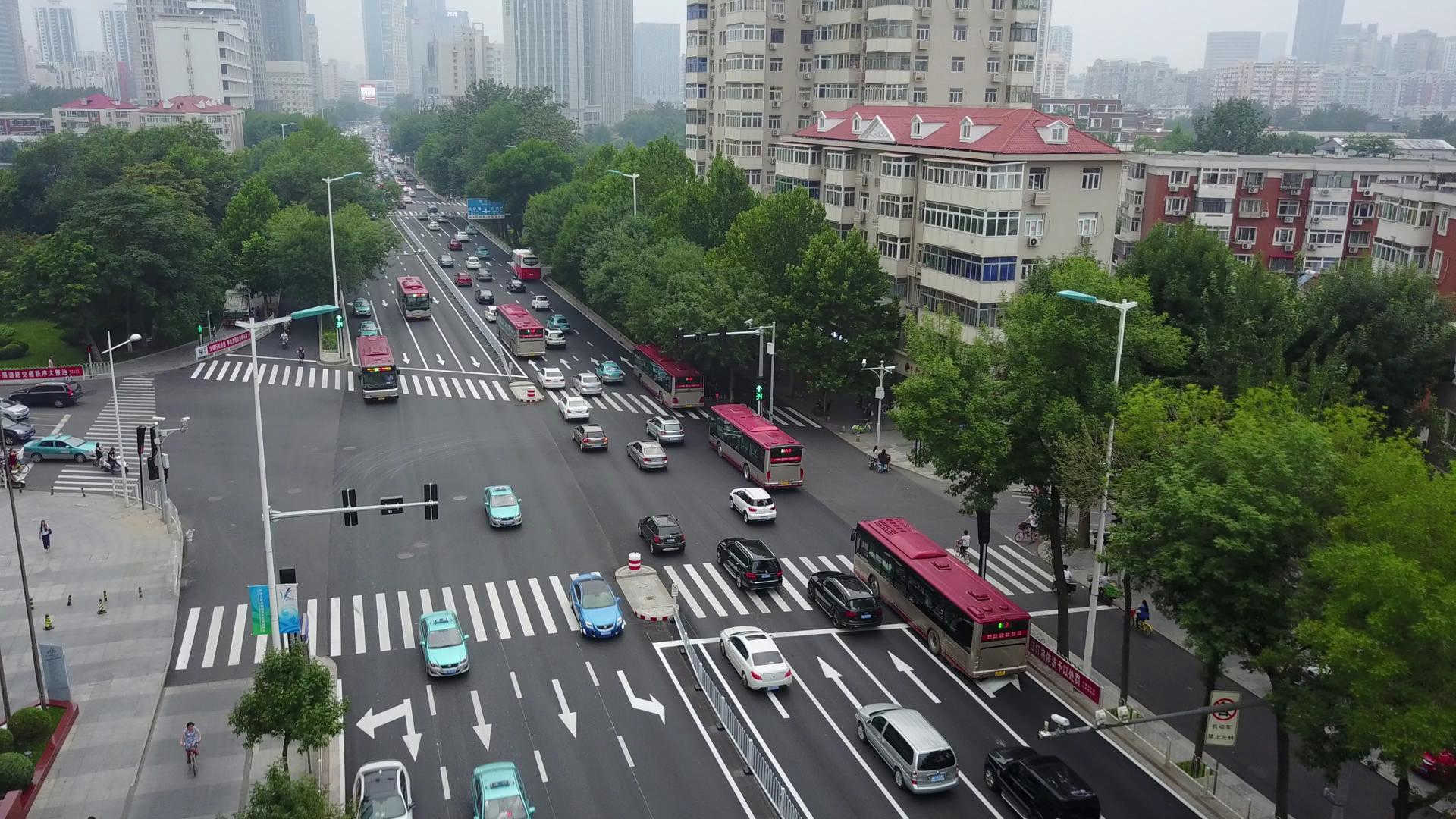}}
        \vspace{0.5cm}
      \subfloat[SIMDive multiplication and division (PSNR = 29.3)\label{sub_fig7}]{%
        \includegraphics[width=0.47\textwidth]{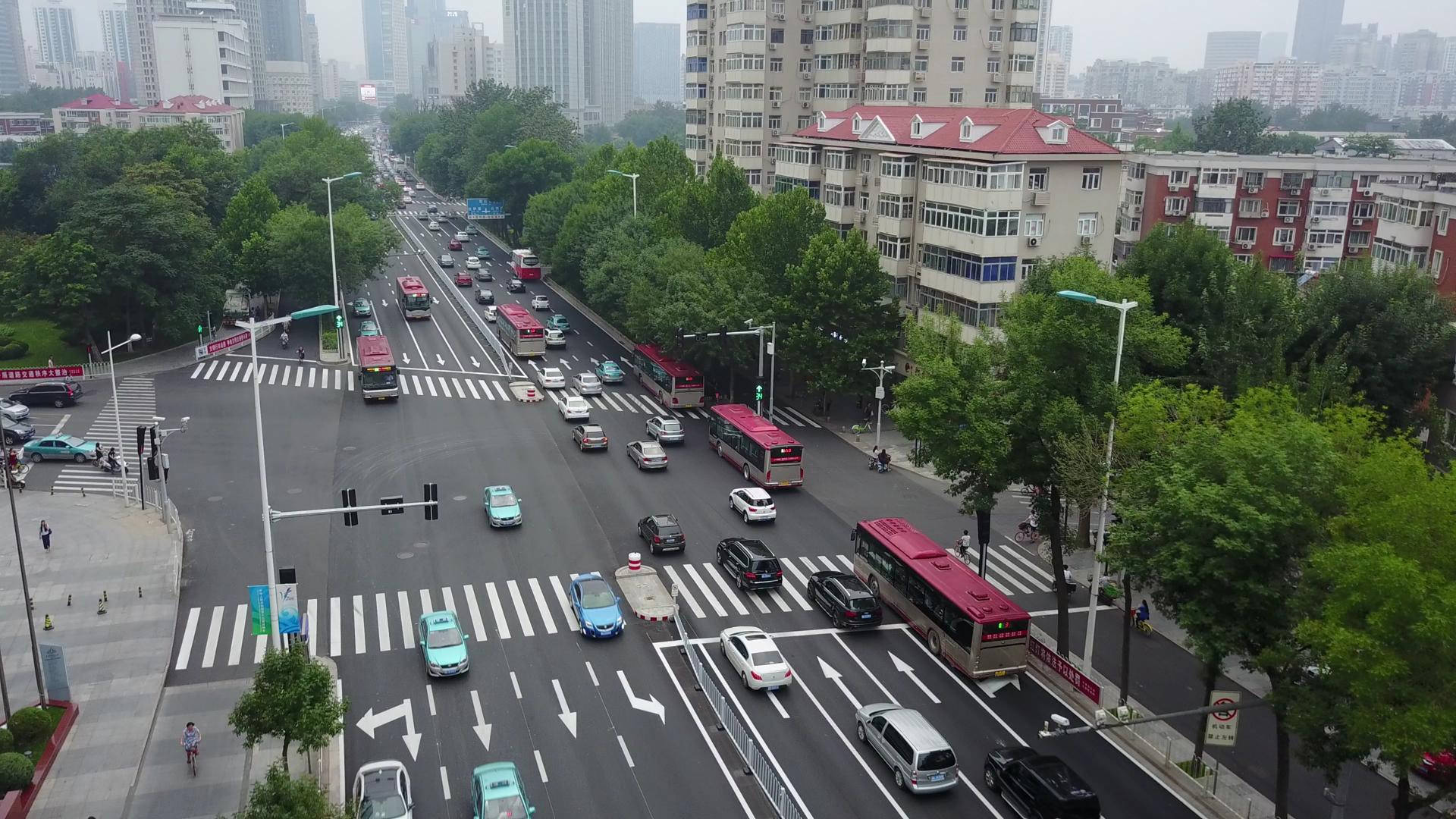}}
            \hfill
      \subfloat[DRUM-6 multiplication and AAXD-8/4 division (PSNR = 24.4)\label{sub_fig8}]{%
        \includegraphics[width=0.47\textwidth]{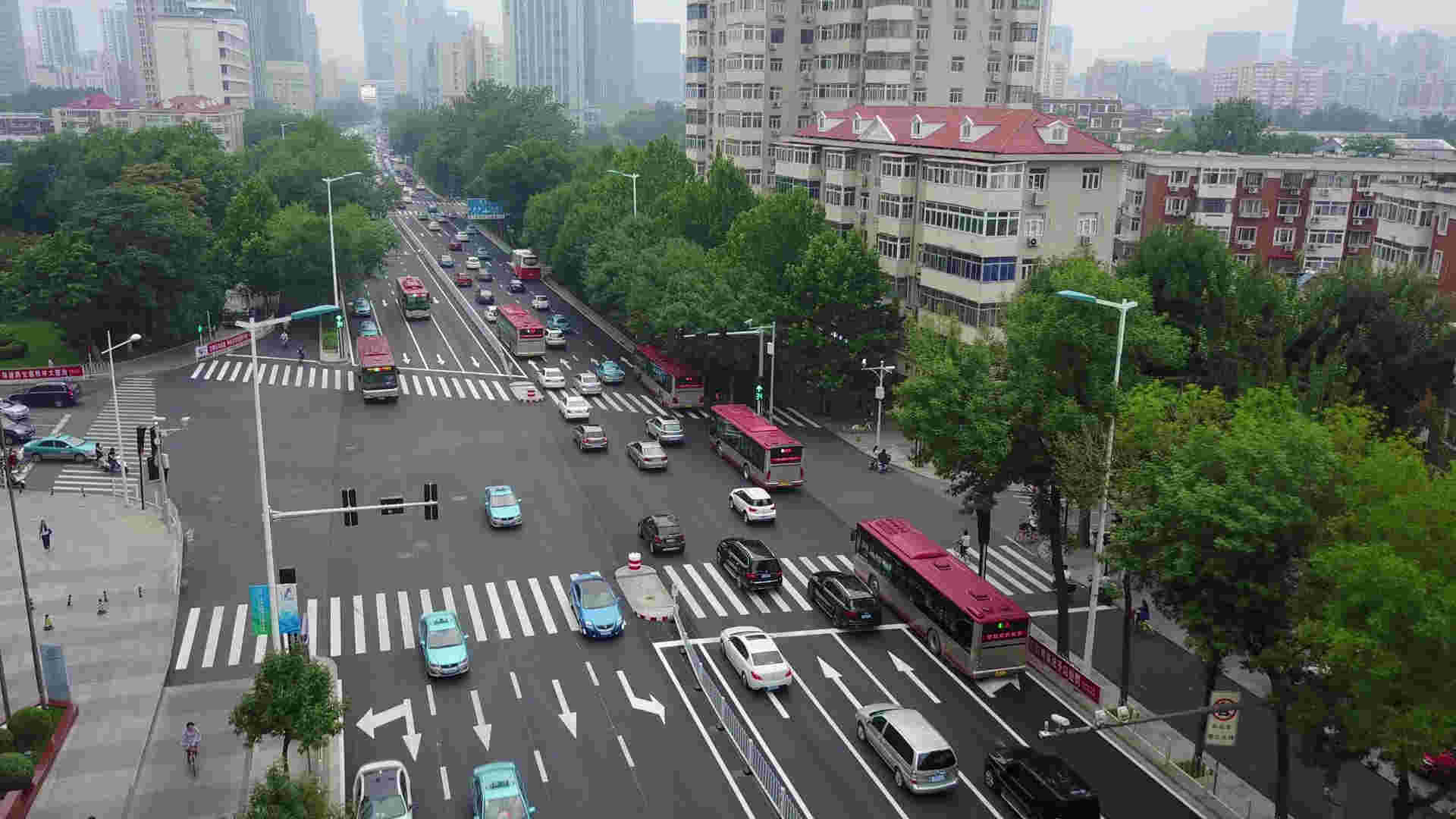}}
 \caption{
 Comparison of JPEG compression on aerial images with accurate and different approximate multipliers and dividers (16-bit).}\label{fig:jpeg_images}
  \vspace{0.1cm}
\end{figure*} 
\vspace{0.1cm}

\begin{figure*}[!h]
   \centering
  \subfloat[Accurate multiplication and division (baseline) = 100\%\label{sub_fig1}]{%
      \adjustbox{trim={.0\width} {.08\height} {0.01\width} {.15\height},clip} 
      {\includegraphics[width=0.47\textwidth]{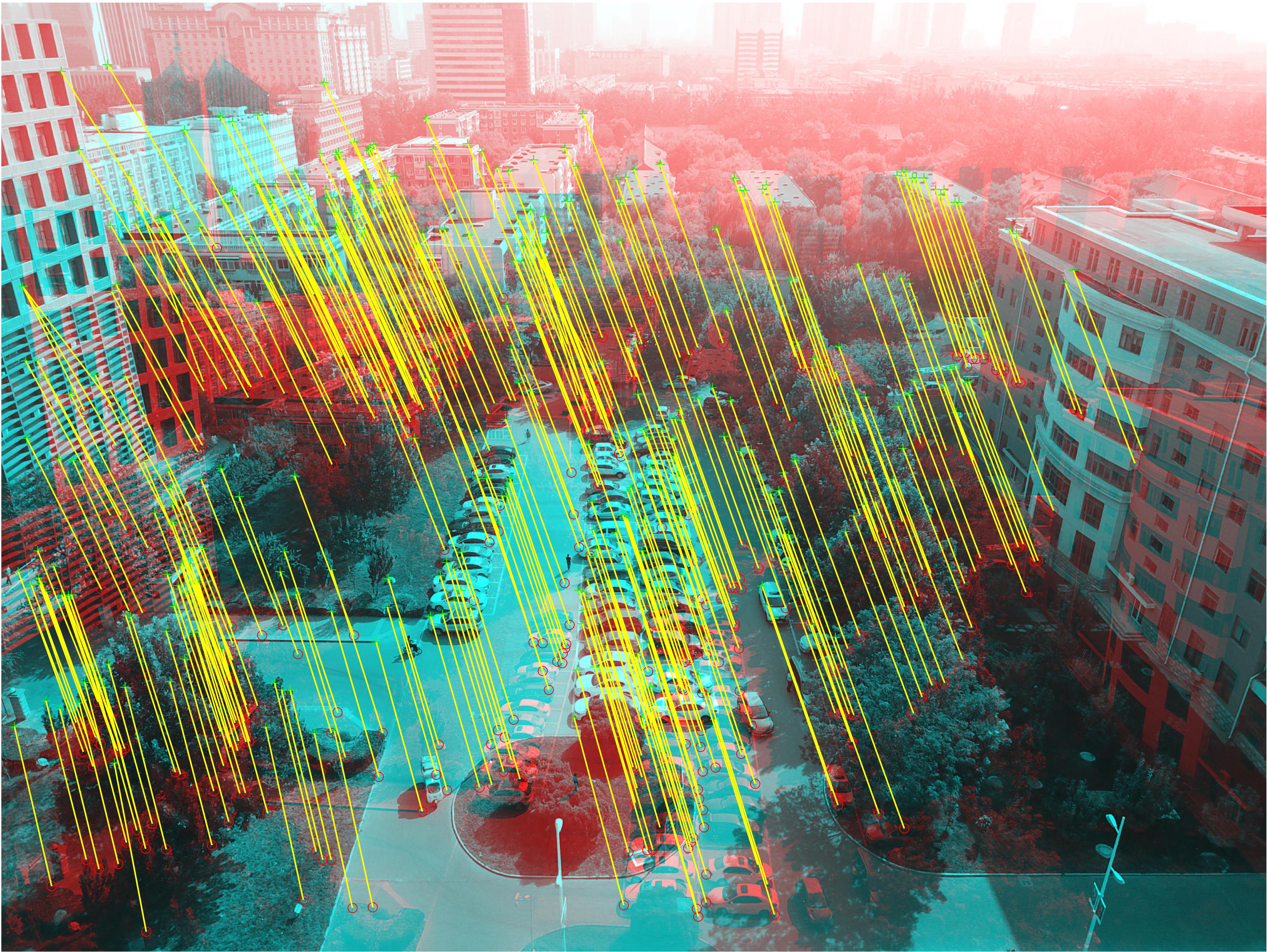}}}
    \hfill
  \subfloat[\pname~multiplier-10 and divider-9 = 94\%\label{sub_fig2}]{%
        \adjustbox{trim={.0\width} {.08\height} {0.01\width} {.15\height},clip}
        {\includegraphics[width=0.47\textwidth]{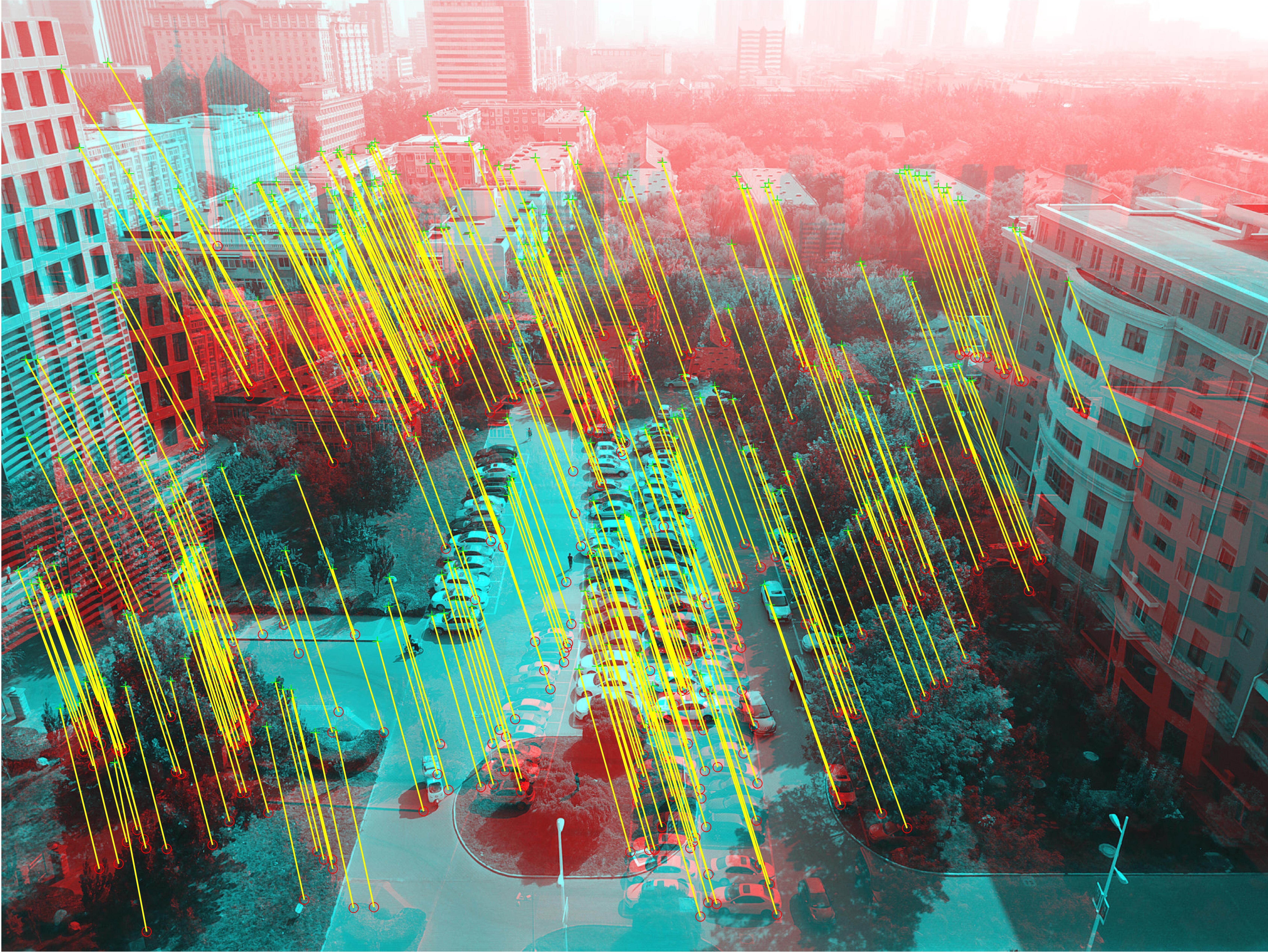}}}
        \vspace{0.3cm}
      \subfloat[SIMDive multiplication and division = 97\%\label{sub_fig3}]{%
        \adjustbox{trim={.0\width} {.08\height} {0.01\width} {.15\height},clip}
        {\includegraphics[width=0.47\textwidth]{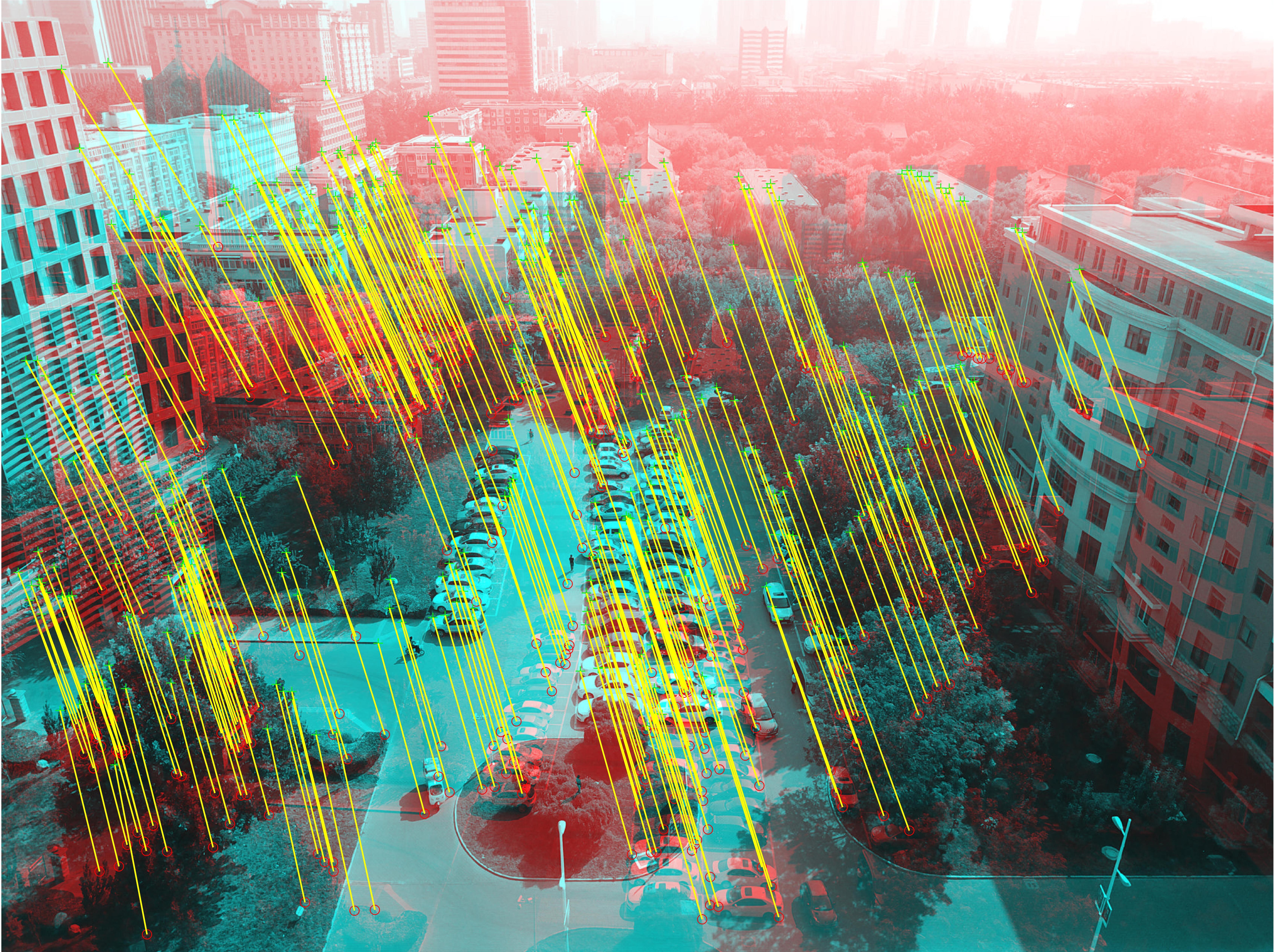}}}
            \hfill
      \subfloat[DRUM-6 multiplication and AAXD-8/4 division =  83\%\label{sub_fig4}]{%
        \adjustbox{trim={.0\width} {.08\height} {0.01\width} {.15\height},clip}
        {\includegraphics[width=0.47\textwidth]{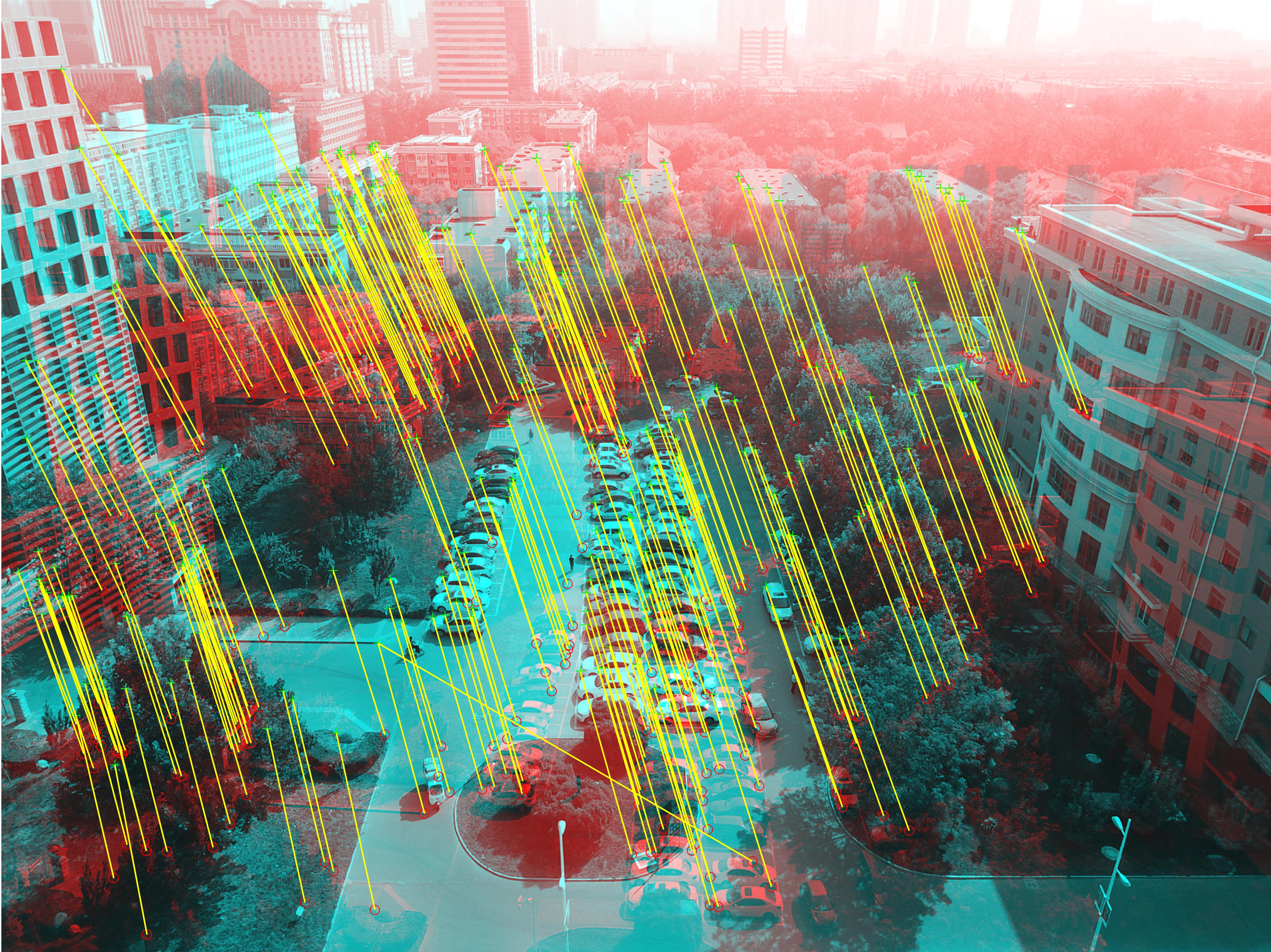}}}
 \caption{
Tracking via Harris Corner Detection: changes in Harris score range also enables detection of new vectors in the threshold- based selection of tracking algorithm in MATLAB. Average of false positive vectors is less than 5\%.
}\label{fig:harris_images}
  \vspace{-0.3cm}
\end{figure*}

\end{itemize}

 \vspace{-0.1cm}

\subsection{Evaluation of \pname~in three multi-kernel applications}
The efficacy of \pname~over accurate and SoA multipliers\footnote{Herein the SISD mode of SIMDive has been analyzed. Detailed comparison of SIMD architectures with pipelining\;for\;multi$\mbox{-}$kernel applications and resolving instruction dependencies are targeted as interesting future works.}
and dividers has been also appraised by deploying them in the end-to-end implementation of three multi-kernel applications. The application, shown in figures \ref{fig:pan_tompkins}, \ref{fig:jpeg}, and \ref{fig:harris} include JPEG compression, heart-beat detection through Pan-Tompkins algorithm, and HCD
(the corners are further employed to generate movement vectors, which is used in object tracking programs).

\noindent\ul{Hardware implementation and performance analysis of applications}: the source-code of applications is synthesized with Xilinx Vivado for the usually-adopted kernel configuration of 16-bit. The implementation of Pan-Tompkins algorithm is adopted from \cite{10.1145/3316781.3317933}. The base of JPEG compression is adopted from AxBench \cite{7755728} and further optimized for a resource-efficient implementation on FPGA, by  e.g., transforming 2D-DCT calculations to the butterfly-based 1D-DCT approach \cite{7755728, 10.1117/12.2006174}. We have developed both  JPEG and HCD in C++
and synthesized them through Vivado High-Level Synthesis (HLS) and
disabled DSPs. HLS has two key advantages: first, it facilitates applying various directives, in a system-level implementation as well as the process for generating different, customized configurations. Second, it simplifies the high-level behavioral evaluation of the entire design.
To efficiently reflect the optimizations of \pname~in the final HDL design and to overcome the resource gap between HLS generated and HDL, we have employed a three-step approach. In the first step we have coded each of accurate multiplication and division functions in the applications. This scheme has facilitated replacing each function with its optimized \pname~versions, later in step 3. Second, the compiler has been forced, via HLS inline pragmas, to generate an independent HDL file for each of multiplication and division functions. In the third step, the HDL description of the respective functions are replaced by HDL-optimized versions of \pname~modules.
Finally, for the end-to-end performance analysis, the HLS-synthesized designs for all accurate and approximate applications have been further passed to the downstream implementation phase, placed and routed on Virtex-7.

\begin{figure*}[!t] 
 \centering
 \includegraphics[width=\textwidth]{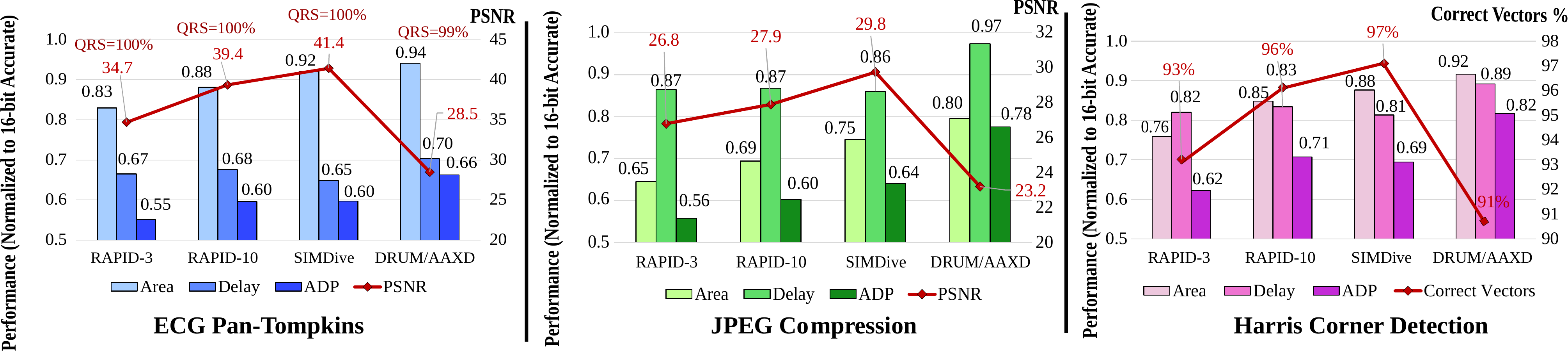}
 \caption{
End-to-end performance of applications utilizing \pname~multiplier and divider, compared to accurate and SoA approximate designs.
}
 \label{fig:RAPID_ApplicationLevel} 
 \vspace{-0.2cm}
\end{figure*}

\begin{figure*}[!t] 
 \centering
 \includegraphics[width=\textwidth]{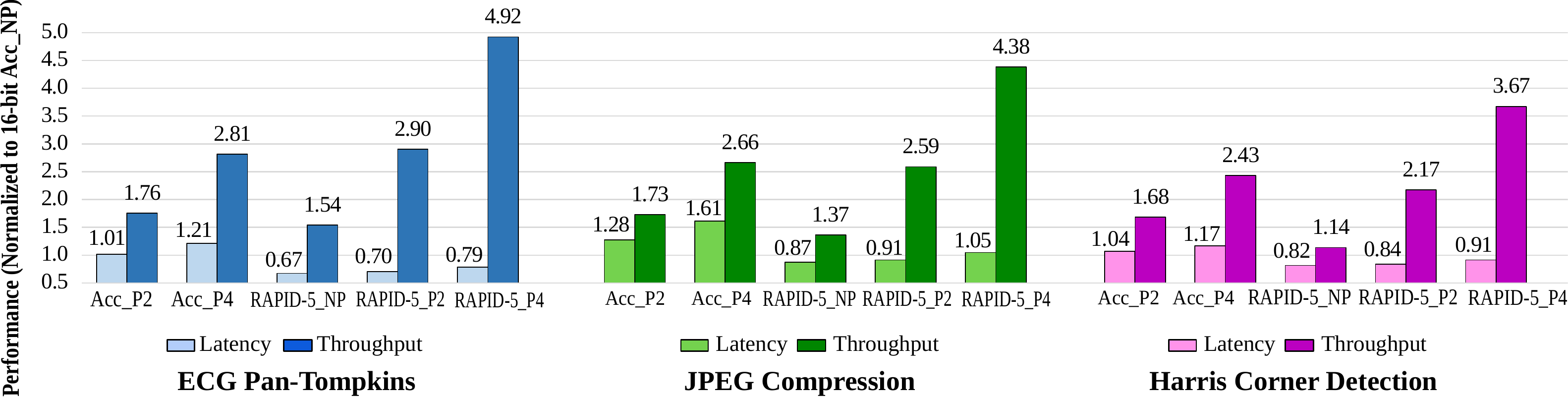}
 \caption{
The end-to-end latency and throughput of applications utilizing \pname~multiplier and divider, compared to accurate IP counterparts, in pipelined and non-pipelined format.
}
 \label{fig:Pipeline_vs_NP_bar} 
 \vspace{-0.2cm}
\end{figure*}

\begin{figure}[!t] 
 \centering
 \includegraphics[width=0.5\textwidth]{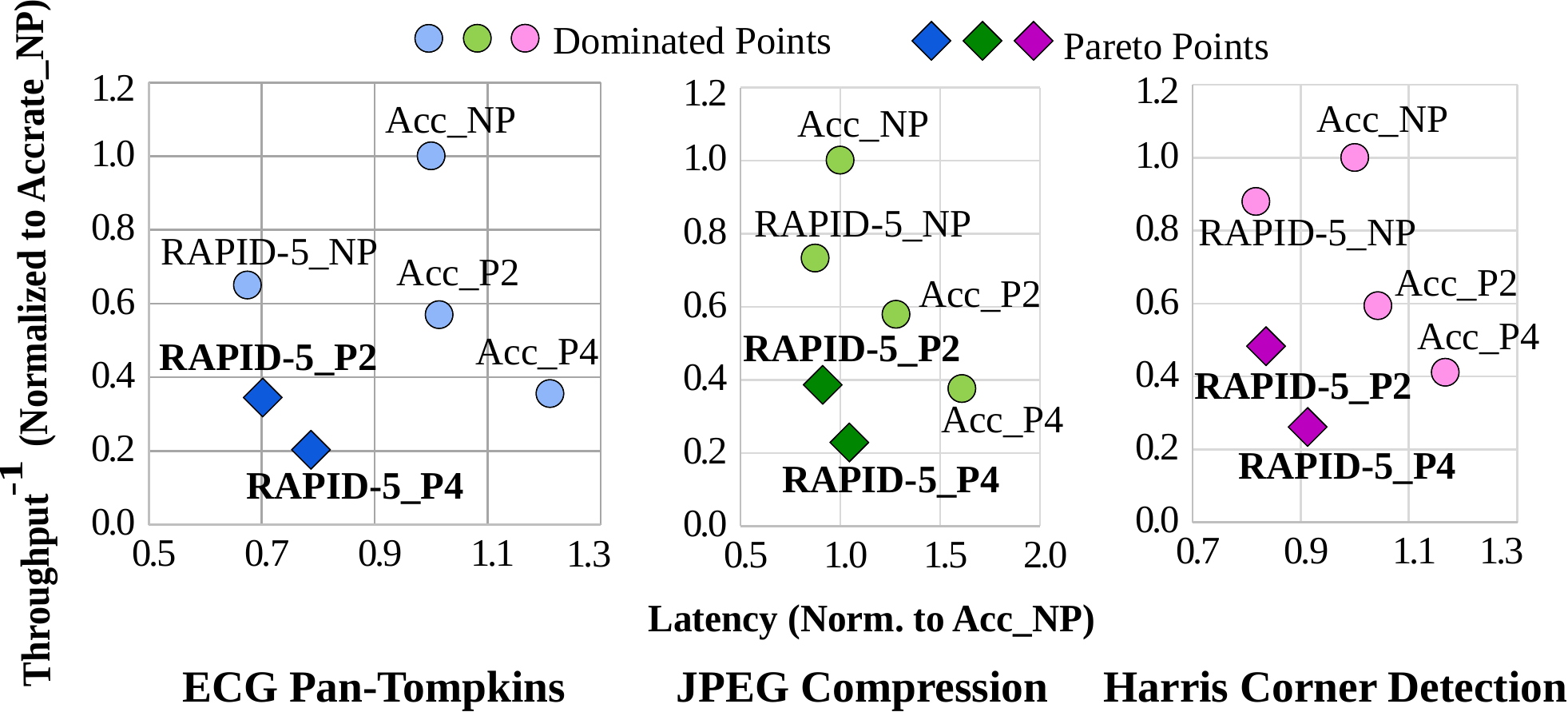}
 \caption{
Comparing the latency and throughput trade-off in applications exploiting pipelined and non-pipelined version of \pname~and accurate multiplier and divider.
}
 \label{fig:Pipeline_vs_NP_Pareto} 
 \vspace{-0.4cm}
\end{figure}

\noindent\ul{QoR analysis on real-world benchmarks}: for assessing the end-to-end accuracy of Pan-Tompkins, we measured the QRS and Peak Signal to Noise Ratio (PSNR) for 30k ECG samples from the MIT-BIH database \cite{b6ea16e20e8d49968b728e3b6db5fcf3} in MATLAB.
For quality measurements on JPEG-compressed images PSNR is used, while the percentage of correct vectors is considered as the application-level metric for HCD. In fact, similar to \cite{9104687}, the extracted corners from HCD algorithm are passed to MATLAB for generating the motion/terrain vectors\footnote{In\,\cite{9104687},\,only DCT \textit{adders}\, are approximated\;(DCT is used as a pre-processing compression step in UAV programs).}. The accuracy changes in these applications are measured by conducting analysis on 50 images from three aerial imagery datasets \cite{10.1007/978-3-319-46448-0_27, 10.1007/978-3-030-66823-5_44, LYU2020108}. To remain inline with industrial standards, we refrained from applying approximations to \textit{zigzag} and \textit{Huffman} kernels in JPEG which implement a re-arrangement and encoding scheme, respectively. Moreover, the \textit{corner selection via non-maximum suppression} in HCD have also been remained accurate, as it is mainly comprised of comparison operations and has moderately low resource footprint. 
In this article we have considered the PSNR of, at least, 28 dB for ECG and JPEG compression and 90\% correct vectors for HCD application (reported to be an acceptable confidence level for moving object tracking \cite{9104687}). Please note, although a relatively low PSNR of 11-19 dB, and thereby, 95-100\% detection ratio has been allowed in XBioSip \cite{10.1145/3316781.3317933}, we adhere to a high PSNR in this article, that is also expected to render 100\% detection accuracy \cite{10.1145/3316781.3317933}.

\noindent\ul{Post-implementation results}: 
Fig. \ref{fig:RAPID_ApplicationLevel} shows the obtained area, delay, and Area-Delay-Product (ADP) improvements of approximate designs, in the three applications\footnote{Comprehensive energy analysis for multi-kernel programs on large image/ECG datasets is for follow$\mbox{-}$up track}. The comparison is among three configurations: adopting with \pname~multipliers and dividers, adopting SoA SIMDive multiplier and divider, or dynamically truncated designs, i.e., DRUM-6 multiplier
together with AAXD-8/4 divider.
The following inferences are highlighted on the efficacy of \pname~designs, after appraising the end-to-end performance gain:

First, referring to Fig. \ref{fig:jpeg_images} and Fig. \ref{fig:harris_images} the quality drop by designs having a near-unbiased error characteristic (both \pname~and SIMDive) is smaller than when truncated counterparts are adopted.
As discussed before, the biased error of truncated designs, i.e., DRUM and AAXD, results in accumulation of errors in consecutive kernels.
In fact, our profiling has revealed that the near-zero biased errors of the \pname~multiplier and divider have been able to cancel out each other in consecutive operations/kernels and prevent a drastic error accumulation in the aggregation-based (mostly Add/Mul) structure of kernels. Such observations also corroborate \cite{8863138, 9186830} in that error-bias feature plays a pivotal role in approximation of consecutive kernels having an aggregation-based structure, e.g., neural networks.
Second, the negligible accuracy loss after deployment of \pname~is partly due to the small average and peak error of the proposed multiplier and divider, as compared to high peak error in DRUM/AAXD, (up to 100\% in hundreds of input cases). In fact, the analysis shows that a high peak error is also reflected as incorrect vectors in the HCD application (especially due to the presence of division in the last stage of the HCD algorithm). This also holds true for false positive heartbeat (QRS peaks), as by utilizing both DRUM and AAXD the detection accuracy of heartbeat has dropped by $\sim$1\%.  
Second, \pname-configured applications also have better area- and ADP- gains, compared to both SIMDive and truncated counterparts (see Fig. \ref{fig:RAPID_ApplicationLevel}). SIMDive on the other hand has marginally better end-to-end latency (by at most 3\%), since it's error-reduction circuitry is customized for performance-efficiency by directly configuring LUTs. Nonetheless, the difference of QoR and performance-gain between \pname~and SIMDive is not significant at application-level.

To compare the throughput of applications in pipelined and non-pipelined configuration, we have also deployed the 2- and 4-stage pipelined versions of the \pname~multiplier and divider (along with their accurate counterparts). For a fair comparison, we have avoided user-specified optimizations such as \textit{function pipelining} pragmas (e.g., on matrix multiplication) and the applications are also implemented on the basis of streaming approach. Fig. \ref{fig:Pipeline_vs_NP_bar} compares the end-to-end latency and throughput of both pipelined and non-pipelined configurations. Please note, the area difference by only replacing the non-pipelined \pname~multipliers and dividers with their pipelined versions is not significant (see Fig. \ref{fig:RAPID_ApplicationLevel}). The throughput is estimated as the inverse of the clock period for the applications, especially as they are constantly fed with bulk of data. As can be seen in Fig. \ref{fig:Pipeline_vs_NP_bar}, the throughput of the applications increases after applying the pipelining, but this comes with the cost of increase in the end-to-end latency. Nevertheless, the latency overhead of converting \pname\_P2 configuration to \pname\_P4 is less than of the overhead from converting Acc\_P2 to Acc\_P4. Similar observation also holds true for the improvement in the throughput. Moreover, application configurations having the \pname\_P2 reach smaller latency along with the higher throughput than when incorporating Acc\_NP or Acc\_P2.
Finally, Fig. \ref{fig:Pipeline_vs_NP_Pareto} illustrates the Pareto points in the trade-off between Latency and throughput for pipelined an non-pipelined architectures. As can be observed, \pname\_P2 and \pname\_P4 render the Pareto points having better trade-off in latency and throughput than other configurations.

\textit{Discussion}: it should be highlighted that although more complicated object tracking or heart arrhythmia programs utilize machine-learning techniques for feature extraction, continuously offloading the complete video stream (for the former) or patient bio-signal data (for the latter) to the insecure/untrustworthy network, or process/store it on the third-party cloud can pose performance bottleneck for a real-time processing and deplete the battery in a short interval.
Therefore, enabling extraction of some features at the edge is highly desired. 

 \section{Conclusions and Future Work} \label{sec:conclusion}

In this paper, we have proposed \pname, the first fine-grain pipelining architecture for approximate multiplier and divider. The proposed error-reduction schemes of \pname~enable 99-99.4\% accuracy with smaller cost, compared to the existing approaches. In particular, pipelined \pname~ multipliers (dividers) enable up to 3.3$\times$ (8.1$\times$) higher throughput, 2.3$\times$ (6.8$\times$) higher throughput/Watt, and 56\% (36\%) savings of LUTs, over pipelined accurate IPs.
The end-to-end evaluations of \pname~in heartbeat detection, JPEG compression, and Harris corner detection demonstrate up to
35\%, 33\%, and 45\% improvements in area, latency, and Area-Delay-Product, respectively, over accurate configuration, with negligible loss in QoR.
\pname~pipelined designs are interesting candidates to speed up the execution of a wide domain of stream-based applications that are constantly fed with a bulk of data. 
35\%, 33\%, and 45\% improvements in area, latency, and Area-Delay-Product (ADP), respectively, over accurate kernels, with negligible loss in QoR.

For future work, we intend to assess the efficacy of the \pname~pipeline mode in different application domains, e.g., Neural networks which offers both SIMD and pipelining opportunities. The challenge is resolving data dependencies in consecutive instructions. Such challenges are usually partially addressed through out-of-order execution
in processors. However, this technique cannot fully utilize the pipelining potentials. Therefore, we target providing specialized versions of the pipelined multiplier and divider, which will be able to support internal data forwarding \cite{8715045, 8877443} and able to resolve data dependencies. It should be noted that bypassing would be faster and posed with smaller overhead, when implemented through an intra-unit granularity.

Furthermore, we plan to design an approximate Arithmetic Logic Unit (ALU) and assess its applicability in the data-path of soft processors such as RISC-V. In fact, \pname~bears a great potential to be deployed in the mantissa multiplier/divider which consume more than 95\% of the total area and power in the floating point unit (in which division latency is up to 35$\times$ of addition operation) \cite{8493590, 8316989}. Recently, this track has attracted noticeable attention, especially due to the ever-growing usage of 3D computer graphics \cite{9187837, 9374478}.

\section*{Acknowledgement}
This research is co-funded by the projects \href{https://cfaed.tu-dresden.de/PD-Project-REAP}{\textcolor{dark_blue}{\textit{X-ReAp: Cross(X)-Layer Runtime Reconfigurable Approximate Architecture}}} (Number 380524764), funded by the German research foundation \href{https://gepris.dfg.de/gepris/projekt/380524764?context=projekt&task=showDetail&id=380524764&}{\textcolor{dark_blue}{\textit{Deutsche Forschungsgemeinschaft (DFG)}}} and  \href{https://cfaed.tu-dresden.de/PD-Project-REAP}{\textcolor{dark_blue}{\textit{Re-learning: Self-learning and flexible electronics through inherent component reconfiguration}}} (Number 100382146), funded by the \href{https://ec.europa.eu/esf/home.jsp}{\textcolor{dark_blue}{\textit{European Social Fund (ESF)}}}.


\printbibliography

\newpage

\begin{IEEEbiography}[{\includegraphics[width=1in,height=1.25in,clip,keepaspectratio]{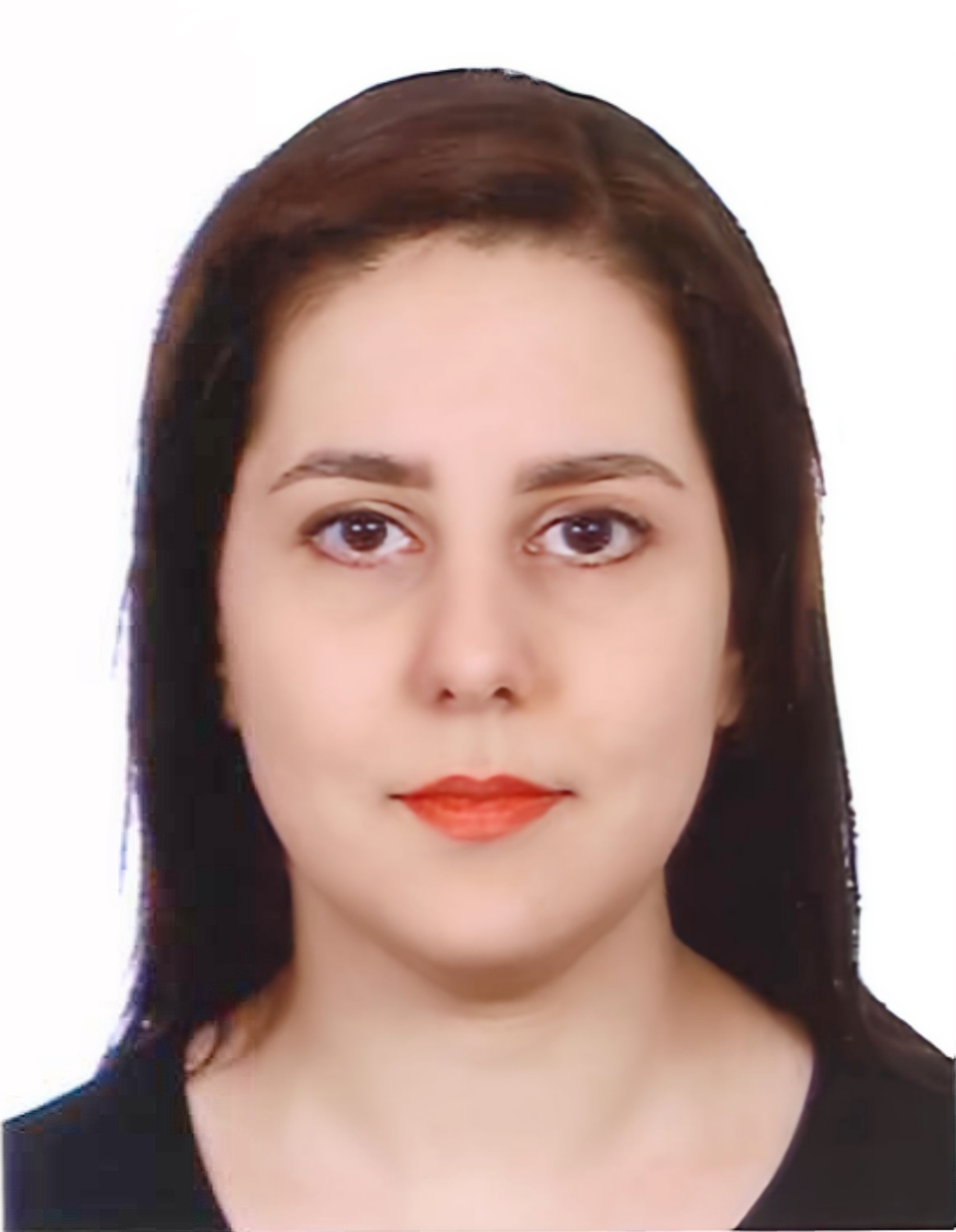}}]{Zahra Ebrahimi} received her B.Sc. and M.Sc. degrees in computer engineering at Sharif University of Technology (SUT), Iran, in 2014 and 2016, respectively. Meanwhile, she was also a Research Assistant at the Data Storage, Networks, and Processing Laboratory, at SUT. She started her Ph.D. at the Center for Advancing Electronics Dresden (cfaed), Technische Universität Dresden, Germany, in 2018. Her research interests include approximate computing, reconfigurable accelerator design, and energy-efficient edge computing.
\end{IEEEbiography}

\begin{IEEEbiography}[{\includegraphics[width=1in,height=1.25in,clip,keepaspectratio]{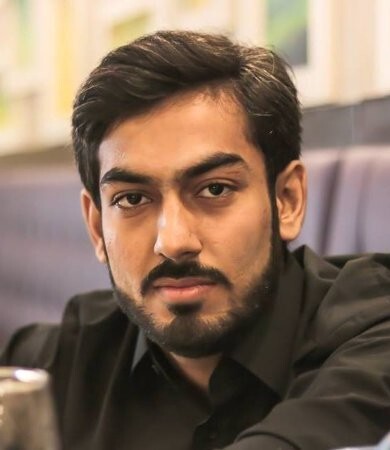}}]{Muhammad Zaid} received his B.Sc. degree in Electrical Engineering from National University of Sciences and Technology (NUST), Islamabad, Pakistan, in 2017. He is currently pursuing the M.Sc. in Nanoelectronic Systems at Technische Universität Dresden. His research interest includes HW/SW co-design of embedded AI systems.
\end{IEEEbiography}

\begin{IEEEbiography}[{\includegraphics[width=1in,height=1.25in,clip,keepaspectratio]{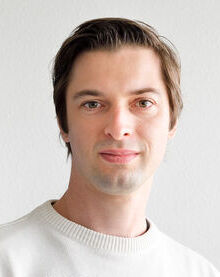}}]{Mark Wijtvliet} received his MS.c. and Ph.D. degrees at the Electronic Systems group at Eindhoven university of technology in 2011 and 2020, respectively. His Ph.D. topic focused at energy efficient reconfigurable processor architectures. Afterwards, he worked as a post-doctoral researcher at TU Dresden at the processor design chair in the field of hardware security. Currently he works as a researcher at ASMPT in the Netherlands. His interests include reconfigurable hardware, hardware security, processor optimization, chip design, and space applications.
\end{IEEEbiography}

\begin{IEEEbiography}[{\includegraphics[width=1in,height=1.25in,clip,keepaspectratio]{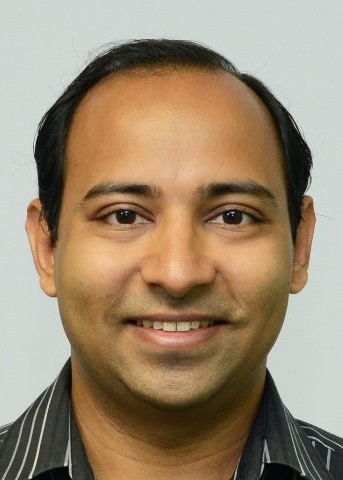}}]{Akash Kumar} (SM’13) received the joint Ph.D. degree in electrical engineering and embedded systems from the Eindhoven University of Technology, Eindhoven, The Netherlands, and the National University of Singapore (NUS), Singapore, in 2009. From 2009 to 2015, he was with NUS. He is currently a Professor with Technische Universität Dresden, Dresden, Germany, where he is directing the Chair for Processor Design. His current research interests include the design, analysis, and resource management of low-power and fault-tolerant embedded multiprocessor systems.
\end{IEEEbiography}

\end{document}